\documentclass[lettersize,journal]{IEEEtran}
\usepackage{tabularx}
\usepackage{amsmath,amsfonts}
\usepackage{amsmath}
\usepackage{mathrsfs} 
\usepackage{algorithm}
\usepackage{algorithmic}
\usepackage{array}
\usepackage{booktabs}
\usepackage{graphicx,latexsym,amsmath,epsfig,latexsym}
\usepackage[caption=false,font=normalsize,labelfont=sf,textfont=sf]{subfig}
\usepackage{cite}
\usepackage{makecell}
\usepackage{caption}
\usepackage{xcite}
\usepackage{multirow}
\usepackage{color}
\usepackage{xcolor}
\usepackage{textcomp}
\usepackage[T1]{fontenc}
\usepackage{stfloats}
\usepackage{setspace}
\usepackage{url}
\usepackage{tabularx}
\usepackage{multirow}
\usepackage{xcolor}
\usepackage{array}
\usepackage{verbatim}
\usepackage{graphicx}
\usepackage{hyperref}
\hypersetup{
    colorlinks=true,
    linkcolor=green,
    citecolor=green,
    urlcolor=magenta
}
\def\BibTeX{{\rm B\kern-.05em{\sc i\kern-.025em b}\kern-.08em
    T\kern-.1667em\lower.7ex\hbox{E}\kern-.125emX}}
\usepackage{balance}
\begin{document}
\title{Aerial IRS Deployment-Aided Secure Computation Offloading Against DISCO Jamming Attacks}
\author{
		
	Minghui Min,~\IEEEmembership{Senior Member,~IEEE}, Peng Zhang, Jiayang Xiao, Ruixin Yang,~\IEEEmembership{Member,~IEEE}, Shiyin Li,\\  Huan Huang,~\IEEEmembership{Member,~IEEE}, Hongliang Zhang,~\IEEEmembership{Member,~IEEE},
	 Zhu Han,~\IEEEmembership{Fellow,~IEEE}
	
	\thanks{ \emph{Corresponding author: Shiyin Li}. Email: lishiyin@cumt.edu.cn.}
	\IEEEcompsocitemizethanks{
		\IEEEcompsocthanksitem
		Minghui Min is with the School of Information and Control Engineering, China University of Mining and Technology, Xuzhou 221116, China, and also with the College of Computing and Data Science, Nanyang Technological University, Singapore 639798. E-mail: minmh@cumt.edu.cn.
		\IEEEcompsocthanksitem
		Peng Zhang, Ruixin Yang, and Shiyin Li are with the School of Information and Control Engineering, China University of Mining and Technology, Xuzhou 221116, China. E-mails: zhang\_peng@cumt.edu.cn, ray.young@cumt.edu.cn, lishiyin@cumt.edu.cn.
		\IEEEcompsocthanksitem
		Jiayang Xiao is with the Ira A. Fulton Schools of Engineering, Arizona State University, Tempe, AZ 85287, USA. E-mail: jxiao123@asu.edu.
		\IEEEcompsocthanksitem Huan Huang is with the School of Electronic and Information Engineering, Soochow University, Suzhou 215006, China. E-mail: hhuang1799@gmail.com.
\IEEEcompsocthanksitem Hongliang Zhang is with the State Key Laboratory of Photonics and Communications, School of Electronics, Peking University, Beijing, 100871, China. E-mail: hongliang.zhang92@gmail.com.
		\IEEEcompsocthanksitem Zhu Han is with the Department of Electrical and Computer Engineering at the University of Houston, Houston, TX 77004 USA, and also with the Department of Computer Science and Engineering, Kyung Hee University, Seoul, South Korea, 446-701. E-mail: hanzhu22@gmail.com.
	}
}

\maketitle

\begin{abstract}
With the rapid growth of Multi-access Edge Computing (MEC), secure and efficient computation offloading from user equipment (UEs) to edge access points (APs) is critical. However, DISCO intelligent reflective surface-based fully-passive jammers (DIRS-based FPJs) use random time-varying phase shifts to launch DISCO jamming attacks, disrupting offloading performance. This paper leverages an aerial intelligent reflective surface (AIRS) to enable secure computation offloading against DISCO jamming by jointly optimizing offloading ratios, AIRS phase shifts and deployment. A two-timescale (2Ts) framework is proposed to address the optimization challenge caused by the distinct update frequencies of different strategies. Specifically, AIRS deployment is adjusted on a long timescale to boost anti-jamming capability due to the impracticality of frequent physical adjustment, while offloading ratios and phase shifts are optimized on a short timescale to adapt to DIRS-jammed dynamic channel conditions. We propose a dual-agent deep reinforcement learning (DRL)-based AIRS deployment-aided secure computation offloading (DDADSO) scheme to maximize the secure offloading utility under DISCO jamming. Simulation results verify that the proposed DDADSO scheme outperforms benchmark schemes, demonstrating the effectiveness of AIRS deployment in improving offloading performance against DISCO jamming attacks.

\end{abstract}


\section{Introduction}
\IEEEPARstart{T}{he} rapid advancements in the sixth-generation (6G) networks and the Internet of Things (IoT) technologies have stimulated a wide range of emerging computation-intensive and latency-sensitive tasks from different applications, e.g., industrial automation, augmented reality (AR), and interactive gaming \cite{luo2025joint}. However, the user equipment (UE), such as smartphones, sensors, and wearable devices, typically possesses limited computation capabilities, making it challenging to support resource-intensive applications and services. In this context, multi-access edge computing (MEC) has emerged as a viable solution by deploying computational capacity and control functions at MEC servers to provide quality of service (QoS) guarantees \cite{wu2025intelligent, bo2025vehicle}.

Recently, the concept of a DISCO intelligent reflective surface (DIRS) as a malicious attacker has been introduced to compromise wireless communication links passively, where an illegitimate IRS employs random time-varying phase shifts, behaving like a ``disco ball'' \cite{HuangHuan2024_5}. These adversarial devices, referred to as DIRS-based fully-passive jammers (DIRS-based FPJs), induce active channel aging (ACA) by reflecting signals in rapidly varying directions, effectively corrupting channel state information (CSI) \cite{HuangFPJ2024}. In MEC uplink systems, this CSI corruption leads to channel deterioration, which weakens the stability of offloading links. Furthermore, CSI mismatch disrupts the effectiveness of the beamforming design for interference suppression, reducing the offloading channel's signal-to-interference-plus-noise ratio (SINR) and increasing the risk of task offloading failure. Consequently, MEC systems under DISCO jamming attacks may fail to meet the real-time QoS requirements of UEs.

To address the above issues, traditional anti-jamming methods (e.g., transmit power control, spread spectrum, and multiple input multiple output (MIMO)) have been employed in wireless networks \cite{Jia2025Surv, Tariq2024RL}. However, fixed transmit power adjustments and basic spread-spectrum methods lack the flexibility to respond dynamically to changing jamming patterns, particularly in environments with DIRS-based FPJs. Besides, adaptive spread-spectrum and MIMO beamforming methods require excessive energy consumption and complex antenna array design, which conflicts with the goal of energy-efficient computation offloading \cite{Jia2025Surv}. These limitations hinder the efficacy of conventional methods in enhancing offloading performance under the DISCO jamming attacks. Fortunately, it has been demonstrated that the induced ACA jamming is not susceptible to mitigation by increasing transmit power alone, and an IRS-aided anti-jamming scheme has been proposed as a solution \cite{Huang2024IRS}. Rather than attempting to directly impact DIRS-based FPJs, the IRS improves the quality of legitimate communication channels and suppresses jamming.

However, fixed-position IRS deployed on buildings can only provide effective service to transceivers on the same side, and lacks the flexibility to adapt to user mobility or time-varying DISCO jamming paths, making it unable to maintain stable line-of-sight (LoS) links for all legitimate users \cite{Zeng2021RIS, Cheng2022RotationLocation}. To address this, recent studies propose aerial IRS (AIRS) by mounting IRS on aerial platforms (e.g., drones, balloons, airships) \cite{Gao2024AIRS, Jiang2024AIRS, Kim2024AIRS}. This design enables panoramic full-angle reflections and flexible 3D positional adjustment, facilitating LoS link establishment and substantially enhancing uplink computation offloading performance compared to fixed IRS \cite{Jiang2024AIRS}. Additionally, UAV-mounted IRS creates strong virtual links and mitigates user privacy and security risks \cite{Kim2024AIRS}. These advantages allow AIRS to establish stable line-of-sight (LoS) links for legitimate offloading and effectively avoid the interference region of DIRS-based FPJs. Consequently, we jointly optimize AIRS deployment and phase shifts to mitigate DISCO jamming attacks.

However, implementing AIRS deployment to enhance offloading performance against DIRS-based FPJs in dynamic MEC systems presents {\em three significant challenges}. First, we have to jointly optimize offloading ratios and phase shifts under varying deployments to meet UEs' real-time QoS requirements \cite{gan2021ris}. The interplay between the AIRS deployment and the AIRS-aided secure offloading policy introduces substantial optimization complexity due to the direct influence of different AIRS deployment designs on offloading channels \cite{AIRS2022Niu}. Second, due to the high cost of physically redeploying AIRS, deployment should occur over a relatively long timescale, with update intervals possibly on the order of several quarters of an hour. However, the execution timeframe for the AIRS-aided offloading strategies is predominantly within several milliseconds and considerably shorter than that of the AIRS redeployment, causing the different update frequencies problem \cite{liu2016delay}. Third, as the DIRS-jammed channels and offloading workloads are unknown in the dynamic MEC systems, we must obtain the optimal AIRS-aided secure offloading strategies adaptively. Traditional optimization methods usually yield suboptimal solutions when applied to the aforementioned complex problems, which limits the performance of AIRS deployment-aided offloading strategies. Furthermore, traditional optimization methods require multiple iterations to converge and may not meet the real-time and dynamic secure offloading requirements. Therefore, we need to design a more efficient AIRS deployment-aided secure computation offloading optimization scheme.

Fortunately, deep reinforcement learning (DRL) algorithms offer a promising solution for navigating complex optimization challenges \cite{Yao2024Anti-Jamming, zhang2024deep }. The state transitions of the dynamic AIRS-aided computation offloading against jamming process can be modeled as a finite-state discrete-time Markov chain \cite{Yao2024Anti-Jamming, wang2022energy}. Consequently, the time-related optimization problem in AIRS-assisted MEC frameworks can be formulated as a Markov decision process (MDP), which can be handled using DRL algorithms. Additionally, advancements in developing a two-timescale (2Ts) framework for the dynamic offloading process have inspired us to formulate a 2Ts framework for the AIRS-deployment aided MEC system \cite{zhang2023two}. The two-timescale problem decomposition ensures the effectiveness of the AIRS deployment-aided secure computation offloading scheme. This differs from a single-timescale joint optimization, which would fail to adapt to DIRS-jammed channels. Additionally, a dual-agent DRL approach was proposed in \cite{wang2022energy} to mitigate timescale discrepancies and interdependencies between edge-cloud offloading schedules and cloud server configurations. Therefore, merging the 2Ts framework with the dual-agent DRL approach is a promising method for acquiring an optimal AIRS-deployment-aided secure computation offloading strategy.

In the proposed framework, our goal is to maximize the secure offloading utility by jointly optimizing the computation offloading ratios of UEs, the phase shifts of AIRS, and AIRS deployment. Acknowledging the varying update frequencies between the AIRS-aided secure offloading and AIRS deployment strategies, we model a 2Ts framework for the AIRS deployment-aided secure MEC system. As the DIRS-jammed channels are time-varying and unknown in dynamic MEC systems, we model the AIRS-aided secure computation offloading process as an MDP in the 2Ts framework and utilize the DRL algorithm to interact with the uncertain environment. We propose a dual-agent DRL-based AIRS deployment-aided secure computation offloading (DDADSO) scheme to acquire the optimal joint deployment and AIRS-aided secure offloading strategy. To reduce the state space dimension and enhance learning stability \cite{Larsen2024VAE}, we design a twin delayed deep deterministic policy gradient with variational autoencoder (TD3-VAE)-based agent for the short-timescale AIRS-aided secure computation offloading problem. Additionally, the deep Q-network (DQN)-based agent learns the AIRS deployment on a long timescale. These two agents achieve the optimization goal of maximizing the secure offloading utility by sharing the reward function and cooperating.

Furthermore, we conduct simulations to demonstrate the influence of power control and the use of AIRS under the DISCO jamming. Additionally, we investigate the MEC system utility under different locations using heatmaps to study the feasibility of AIRS deployment. On this basis, to underscore the performance advantages of the DDADSO scheme, we compare the offloading utility of DDADSO with that of the DICO W/O AD scheme \cite{ wang2022resource} and the DADCO scheme \cite{Shang2022AIRS}. Additionally, we examine how the number of AIRS and DIRS reflecting elements impacts system performance. The simulation results demonstrate that the DDADSO scheme significantly outperforms the benchmarks in terms of offloading utility, effectively enhancing dynamic offloading performance against DISCO jamming.
The core contributions of this paper can be summarized as follows:
\begin{enumerate}
		\item
        \textbf{\textit{DISCO Jamming in MEC Systems:}} We consider a unique DISCO jamming attack in which DIRS acts as a FPJ using random and time-varying phase shifts to severely corrupt the AP's CSI acquisition. This disrupts the effectiveness of the beamforming design intended for interference suppression, thereby reducing the SINR and ultimately increasing the risk of task offloading failures. We implement AIRS to exploit the spatial degrees of freedom and design a favorable AIRS deployment to mitigate the DISCO jamming attacks.
		\item
        \textbf{\textit{Two-timescale Optimization Framework for Secure AIRS-aided Computation Offloading:}} Due to the high cost and practical constraints of physically redeploying AIRS, the deployment is updated on a relatively long timescale. To align with the different update frequencies of the AIRS-aided computation offloading and AIRS deployment strategies, we optimize the phase shifts and offloading ratios on a short timescale, while optimizing the AIRS deployment on a long timescale.
        \item
        \textbf{\textit{Improved Dual-agent DRL-based Secure Computation Offloading Scheme:}} To ensure adaptability in time-varying offloading channels under DISCO jamming attacks, we introduce the DDADSO scheme, an AIRS deployment-aided secure computation offloading scheme based on dual-agent DRL. In the DDADSO scheme, the TD3-VAE and DQN-based agents maximize secure offloading utility by sharing a reward function and cooperating. The VAE encoder is integrated to reduce redundancy in the high-dimensional state space.
		\item
		\textbf{\textit{Performance Evaluations and Analyses:}} Numerical simulations are conducted to show that the SINR and secure utility increase with the joint AIRS-aided computation offloading and AIRS deployment strategies. Moreover, comparison results further show that the proposed DDADSO algorithm outperforms the conventional DRL algorithm and single-timescale optimization algorithm, further indicating its efficiency and robustness against DISCO jamming attacks.
	\end{enumerate}
	
The rest of our paper is structured as follows. Section \ref{RelatedWork} reviews related literature, and Section \ref{SystemModel} presents the system model. Then, Section \ref{DDADSO} presents a DDADSO scheme for learning the joint deployment and offloading strategy against DISCO jamming. Moreover, the results of the simulation experiment and analysis are provided in Section \ref{SimulationResults}.
Finally, we conclude this paper in Section \ref{Conclusion}.

\section{Related Work}\label{RelatedWork}

\subsection{DISCO Jamming Attacks}
DIRS, unlike legitimate IRS, disrupts computation offloading in dynamic MEC environments. Specifically, DISCO jamming induces active channel aging, disabling the channel estimation process and significantly degrading system performance \cite{HuangHuan2024_5}. Unlike traditional jamming, DIRS leverages legitimate signals to actively age channels without requiring extra power or CSI, thereby limiting the effectiveness of classical anti-jamming methods, such as spread spectrum and MIMO interference cancellation \cite{HuangFPJ2024}. Statistical analysis of DIRS-jammed channels shows that they converge to a complex Gaussian distribution. This highlights the complexity of the interference and makes it more difficult to mitigate its impact on offloading reliability \cite{HuangHuan2024_5}. Furthermore, advanced statistical anti-jamming precoding techniques that optimize the signal-to-jamming-plus-noise ratio (SJNR) have shown promise in mitigating DISCO jamming \cite{YuZhouyuan2024_6}.

\subsection{AIRS-aided Physical Layer Security and Secure Transmission Systems}
Physical layer security (PLS) exploits wireless channel characteristics to achieve secure transmission, and IRS has become a key enabler due to its flexible signal modulation capability \cite{Yao2025RISecure, Yao2025RISDRL}. Unlike conventional IRS, AIRS deploys aerial platforms to realize adjustable, mobile signal reflection, providing users with more adaptive coverage and stable LoS links \cite{yang2023joint}. Recent studies have also verified AIRS as a promising solution for enhancing PLS in wireless networks \cite{Gao2024AIRS, Arzykulov2024ARIS}, and UAV-mounted IRS has been applied in secure offloading systems to counter internal eavesdroppers \cite{Kim2024AIRS}. While these studies have improved AIRS-aided communication secrecy, they mostly focus on eavesdropper-related security issues and pay limited attention to secure offloading against specific jamming attackers. Although an IRS-aided anti-jamming scheme exists for DISCO jamming mitigation \cite{Huang2024IRS}, the impact of DISCO jamming on computation offloading performance and dynamic optimization for secure offloading under such scenarios remains under-explored.

\subsection{Optimization in AIRS-aided Secure MEC Systems}
Extensive research has focused on integrating IRS and AIRS into MEC to enhance offloading performance, and it has gradually evolved from traditional performance optimization to secure MEC optimization considering malicious attacks \cite{Jiang2024AIRS, Kim2024AIRS, yang2023joint}. Early studies primarily employed traditional optimization methods to overcome key challenges, such as enhancing energy efficiency \cite{Jiang2024AIRS} and ensuring robust offloading against multiple potential internal eavesdroppers \cite{Kim2024AIRS} for IRS-aided MEC systems. However, these methods rely on accurate channel state information and static system models and are limited in dynamic MEC environments with time-varying channels and workloads. With the advancement of intelligent optimization algorithms, recent studies have adopted DRL to adaptively optimize computation offloading, resource allocation, and secure beamforming \cite{wang2022resource, Zhang2023IRS, Yao2025RISDRL}. Therefore, this paper addresses the dynamic joint optimization of AIRS-aided computation offloading and deployment to mitigate DISCO jamming by means of AI-driven optimization algorithms.

\subsection{DRL Algorithms Used in the 2Ts Optimization}
Traditional optimization methods struggle to handle uncertain jamming attack behavior. However, DRL techniques offer a promising solution for learning and adapting from environmental feedback \cite{Yao2024Anti-Jamming}. Furthermore, the 2Ts framework has been proposed in \cite{wang2022energy, zhang2023two}, enabling effective optimization across varying update frequencies. A 2Ts framework was implemented by combining methods from matching theory and deep reinforcement learning to solve varying timescale optimization problems in \cite{wang2022resource}. Besides, a dual-agent DRL approach was designed in \cite{Yu2021When}, which encompasses a fast-timescale and a slow-timescale learning process, respectively. Consequently, dual-agent DRL emerges as a powerful approach for addressing the complex and multi-timescale optimization challenges inherent in AIRS-aided MEC networks. To enhance computational efficiency by reducing the state space dimension, further works have utilized the VAE to improve the DRL algorithm \cite{Larsen2024VAE, Sun2025GAI}, which inspires us to further improve the algorithm.

\section{System Model}\label{SystemModel}

\begin{figure}[t]
	\setlength{\abovecaptionskip}{0cm}  
	\setlength{\belowcaptionskip}{-0.5cm}  
	\begin{center}
		\includegraphics[height=2.05 in]{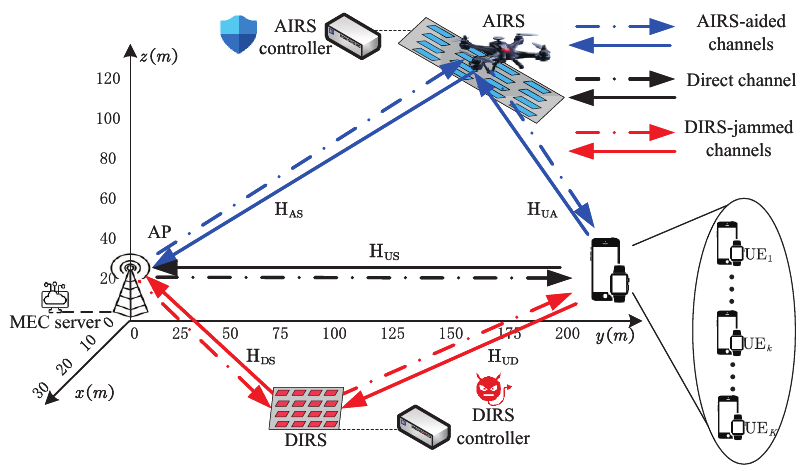}
	\end{center}
	\caption{Illustration of the AIRS-aided MEC system against DISCO jamming attacks. Specifically, we deploy an AIRS to provide secure virtual LoS paths.}
	\label{System Model}
\end{figure}

\subsection{Architecture of AIRS-aided MEC System}
As depicted in Fig. \ref{System Model}, there are $K$ single-antenna UEs denoted as UE$_1$, $\ldots,$ UE$_k$, $\ldots,$ UE$_K$ and an $M$-antenna AP. The offloading channels between the UEs and AP face DISCO jamming attacks because a DIRS is located at $\varUpsilon _D\left( x_D,y_D,z_D \right)$ near the AP.
We deploy an AIRS comprising $N_A$ reflecting elements to assist the computation offloading.
Specifically, AIRS is mounted on a smooth aerial platform, which can hover at a fixed three-dimensional (3D) location $\varUpsilon _A\left( x_A,y_A,z_A \right)$ and provide stable service for a long time \cite{AIRS2022Niu}. Additionally, a smart controller equipped on the AIRS can communicate with the AP through a dedicated backhaul link, allowing them to share CSI information \cite{gan2021ris}. The AIRS is deployed at a suitable location far from the DIRS, and there are two IRS reflections in the UE-AIRS-DIRS-AP cascaded channel. Hence, we consider that the AIRS phase shifts optimization is not affected by the DIRS-based FPJ.
Furthermore, the AIRS on the aerial platform can be deployed at an altitude $z_A\in \left[ z_{A\min},z_{A\max} \right]$, where $z_{A\min}$ represents the lowest altitude set for safety or to avoid obstacles, and $z_{A\max}$ is the highest altitude allowed by regulations \cite{Cheng2022RotationLocation}. Additionally, the AP is located at the origin $\varUpsilon _S\left( 0,0,0 \right)$ and the AIRS surface is parallel to the X-axis-Origin-Y-axis (XOY) plane.

\subsection{DISCO Jamming Attacks} \label{DIRS}
In this paper, the baseband equivalent channels from UE$_k$ to the AIRS, from UE$_k$ to the AP, and from the AIRS to the AP are respectively represented by ${\boldsymbol{h}_{{\text{UA}},k}}\in {{\mathbb{C}}^{1\times N_A}}$, ${\boldsymbol{h}^H_{{\text{US}},k}}\in {{\mathbb{C}}^{ M\times 1}}$, and ${\text{H}_{\text{AS}}}\in {{\mathbb{C}}^{N_A\times M}}$, where $\mathbb{C}^{N_A\times M}$ represents a complex-valued matrix within the dimensions of $N_A \times M$.
Moreover, we denote $\boldsymbol{\Theta } _{A} =\text{diag}\left( \rho _{A,1}e^{j\theta _{A,1}},\ldots,\rho _{A, N_A}e^{j\theta _{A, N_A}} \right)$ as the AIRS reflecting matrix, where $\theta _{A,1},\ldots,\theta _{A, N_A}$ are the AIRS reflecting phase shifts. The amplitudes $\rho _{A,1},\ldots,\rho _{A,N_A}$ are set to 1\cite{bai2020latency, Zhang2023IRS}. Besides, the DIRS reflecting matrix is denoted by $\boldsymbol{\Theta } _{D}=\text{diag}\left( \rho _{D,1}e^{j\theta _{D,1}},\ldots,\rho _{D, N_D}e^{j\theta _{D, N_D}} \right)$. The time-varying DIRS phase shifts $\theta _{D,1},\ldots,\theta _{D,N_D}$ are randomly selected from a discrete set $\left\{ \theta _{D,1},\ldots,\theta _{D,2^{b_D}} \right\}$ with $b_D$ quantization bits. The amplitudes $\rho _{D,1},\ldots,\rho _{D,N_D}$ are related to the corresponding phase shifts $\theta _{D,1},\ldots,\theta _{D,N_D}$ \cite{HuangHuan2024_5}. The DIRS reflecting matrix $\boldsymbol{\Theta } _{D}$ is time-varying for the data transmission (DT) phase and the DIRS with random phase shifts acts like a ``disco ball''. To simplify the presentation, we represent the time-varying DIRS-jammed channel and the AIRS-cascaded channel between UE$_k$ and the AP as $\boldsymbol{h}_{D,k} = \text{H}_{\text{DS}}^{H}\boldsymbol{\Theta } _{D}\boldsymbol{h}_{{\text{UD}},k} $ and $\boldsymbol{h}_{A,k} = \text{H}_{\text{AS}}^{H}\boldsymbol{\Theta } _{A}\boldsymbol{h}_{{\text{UA}},k}  $, respectively. Furthermore, the UE-AP channels, the UE-DIRS channels, the UE-AIRS channels, the DIRS-jammed channels, and the AIRS-cascaded channels for all the $K$ UEs are given by $\text{H}_{\text{US}}^{H}=[ \boldsymbol{h}_{{\text{US}},1}^{H},\ldots,\boldsymbol{h}_{{\text{US}},K}^{H} ] $, $\text{H}_{\text{UD}}^{H}=[ \boldsymbol{h}_{{\text{UD}},1}^{H},\ldots,\boldsymbol{h}_{{\text{UD}},K}^{H} ] $, $\text{H}_{\text{UA}}^{H}=[ \boldsymbol{h}_{{\text{UA}},1}^{H},\ldots,\boldsymbol{h}_{{\text{UA}},K}^{H} ] $, $\text{H}_{D}^{H}=[ \boldsymbol{h}_{D,1}^{H},\ldots,\boldsymbol{h}_{D,K}^{H} ] $, $\text{H}_{A}^{H}=[ \boldsymbol{h}_{A,1}^{H},\ldots,\boldsymbol{h}_{A,K}^{H} ] $.

During the pilot transmission (PT) phase, the CSI for $\text{H}_{\text{US}}$ and $\text{H}_{A}$ can be acquired by AP using current channel estimation methods \cite{YuZhouyuan2024_6}. We utilize the zero-forcing beamforming (ZFBF) method to reduce the inter-user interference \cite{ngo2013energy}, which utilizes the CSI obtained from the PT phase. In particular, the overall beamforming matrix $\mathbf{W}$ is derived by
\begin{equation}\label{H}
	\begin{aligned}
	\mathbf{W}^H &= \left( \left( \text{H}_A+\text{H}_{\text{US}}^{H} \right) \left( \text{H}_A+\text{H}_{\text{US}}^{H} \right) ^H \right) ^{-1}\left( \text{H}_A+\text{H}_{\text{US}}^{H} \right) ^H\\
	&=\left[ \boldsymbol{w}_{1}, \ldots, \boldsymbol{w}_{k}, \ldots, \boldsymbol{w}_K \right] ^H.\\
\end{aligned}
\end{equation}

However, the DIRS-based FPJ remains ``silent'' during the PT phase and the AP cannot obtain the CSI for $\text{H}_{D}$ when designing the overall beamforming matrix $\mathbf{W}$. Moreover, as long as the period of DIRS phase shift variation is close to the time length of the PT phase, the AP cannot acquire any useful knowledge about DIRS-jammed channels through retraining \cite{HuangFPJ2024}. Consequently, the offloading channels are interfered with by the time-varying DIRS-jammed channels.
In MEC systems, this CSI corruption disrupts the effectiveness of beamforming design for interference suppression and leads to channel deterioration that weakens the stability of offloading links.
Based on the previous investigation \cite{Huang2024IRS}, we introduce the legitimate IRS to weaken the strength of the DISCO jamming. Then, the channels during the DT phase between the AP and UE$_k$ are expressed as
\begin{equation}\label{H}
\begin{aligned}
\setlength{\abovedisplayskip}{0.1cm}
\setlength{\belowdisplayskip}{0.1cm}
\boldsymbol{h}_{\text{DT},k}^{}=\boldsymbol{h}_{A,k}+\boldsymbol{h}_{{\text{US}},k}+\boldsymbol{h}_{D,k} .
\end{aligned}
\end{equation}
Then, the SINR $\varGamma_k$ of the offloading channel can be given by
\begin{equation}\label{SJNR}
\varGamma _k=\frac{p_k\left| \boldsymbol{w}_{k}^{H}\left( \boldsymbol{h}_{A,k}+\boldsymbol{h}_{\text{US,}k} \right) \right|^2}{\sum_{i=1}^K{p_i\left| \boldsymbol{w}_{k}^{H}\boldsymbol{h}_{D,i} \right|^2}+\sigma ^2\lVert \boldsymbol{w}_k \rVert ^2},
\end{equation}
where $\sigma^2$ represents the variance of Gaussian white noise and $p_k$ represents the transmit power for UE$_k$.
Hence, the offloading rate $R_k$ between UE$_k$ and the AP is defined as
\begin{equation}\label{Ru}
\setlength{\abovedisplayskip}{0.1cm}
\setlength{\belowdisplayskip}{0.1cm}
R_k=B\log _2\left( 1+\varGamma _k\right),
\end{equation}
where $B$ represents the average bandwidth of all UEs.

\subsection{Channel Model} \label{channel}
The UE-AP channel $\text{H}_{\text{US}}$ and UE-DIRS channel $\text{H}_{\text{UD}}$ are regarded as Rayleigh fading channels \cite{Huang2024IRS}, which is given by
\begin{align}\label{hUS}
\setlength{\abovedisplayskip}{0.1cm}
\setlength{\belowdisplayskip}{0.1cm}
\text{H}_{\text{US}}=\left[ \sqrt{\mathscr{L}_{{\text{US}},1}}\widetilde{\boldsymbol{h}}_{{\text{US}},1},\ldots,\sqrt{\mathscr{L}_{{\text{US}},K}}\widetilde{\boldsymbol{h}}_{{\text{US}},K} \right],
\end{align}
\begin{align}\label{hUD}
\setlength{\abovedisplayskip}{0cm}
\setlength{\belowdisplayskip}{0cm}
\text{H}_{\text{UD}}=\left[ \sqrt{\mathscr{L}_{{\text{UD}},1}}\widetilde{\boldsymbol{h}}_{{\text{UD}},1},\ldots,\sqrt{\mathscr{L}_{{\text{UD}},K}}\widetilde{\boldsymbol{h}}_{{\text{UD}},K} \right],
\end{align}
where $\text{diag}\left\{ \mathscr{L}_{\text{US},1},\cdots ,\mathscr{L}_{\text{US},K} \right\}  $, $\text{diag}\left\{ \mathscr{L}_{\text{UD},1},\cdots ,\mathscr{L}_{\text{UD},K} \right\} $ are the large-scale channel fading coefficients \cite{Huang2024IRS}. $\widetilde{\boldsymbol{h}}_{\text{US}}\sim\mathcal{C N}(0,1)$  and $\widetilde{\boldsymbol{h}}_{\text{UD}}\sim\mathcal{C N}(0,1)$ are assumed to be independent and identically distributed (i.i.d.) complex Gaussian random variables with zero mean and unit variance.

The DIRS-AP channel $\text{H}_{\text{DS}}$ can be given by
\begin{align}\label{hDS}
\setlength{\abovedisplayskip}{0.1cm}
\setlength{\belowdisplayskip}{0.1cm}
\text{H}_{\text{DS}}=\sqrt{\mathscr{L}_{\text{DS}}}\left( \sqrt{\frac{\iota _{\text{DS}}}{\iota _{\text{DS}}+1}}\overline{\text{H}}_{\text{DS}}+\sqrt{\frac{1}{\iota _{\text{DS}}+1}}\widetilde{\text{H}}_{\text{DS}} \right),
\end{align}
where $\iota _{\text{DS}}$ is the Rician factor and $\mathscr{L}_{\text{DS}}$ is the large-scale channel fading coefficient, and $\widetilde{\text{H}}_{\text{DS}}$ is assumed to follow Rayleigh fading \cite{Huang2024IRS, Wu2019IRS}. The deterministic component $\overline{\text{H}}_{\text{DS}}$ is the LoS component, and can be expressed as
\begin{align}\label{hDSLOS}
\setlength{\abovedisplayskip}{0.1cm}
\setlength{\belowdisplayskip}{0.1cm}
\left[ \overline{\text{H}}_{\text{DS}} \right] _{i,j}=e^{-j\frac{2\pi}{\lambda}\left( D_{j}^{i}-D_j \right)},
\end{align}
where $\lambda$ represents the wavelength, $ D_{j}^{i}$, $i=1,\ldots,N_D,j=1,\ldots,M$ is the distance between the $j$-th antenna of the AP and the $i$-th DIRS reflecting element, and $ D_{j}$ is the distance between the $j$-th antenna of the AP and the center of the DIRS.

Furthermore, we optimize the AIRS deployment to provide virtual and secure LoS paths against the DISCO jamming attacks \cite{Shang2022AIRS, Cheng2022RotationLocation}. Hence, UE-AIRS channel $\text{H}_{\text{UA}}$ and AIRS-AP channel $\text{H}_{\text{AS}}$ are assumed as Rician fading channels \cite{Cheng2022RotationLocation, Wu2019IRS}, which are modeled as
\begin{align}\label{hUA1}
\setlength{\abovedisplayskip}{0cm}
\setlength{\belowdisplayskip}{0cm}
\text{H}_{\text{UA}}=\left[ \sqrt{\mathscr{L}_{\text{UA,}1}}\widehat{\text{H}}_{\text{UA,}1},...,\sqrt{\mathscr{L}_{\text{UA,}K}}\widehat{\text{H}}_{\text{UA,}K} \right],
\end{align}
\begin{align}\label{hUA2}
\setlength{\abovedisplayskip}{0cm}
\setlength{\belowdisplayskip}{0cm}
\widehat{\text{H}}_{\text{UA,}k}=\left( \sqrt{\frac{\iota _{\text{UA}}}{\iota _{\text{UA}}+1}}\overline{\text{H}}_{\text{UA,}k}+\sqrt{\frac{1}{\iota _{\text{UA}}+1}}\widetilde{\text{H}}_{\text{UA,}k} \right),
\end{align}
\begin{align}\label{hAS}
\setlength{\abovedisplayskip}{0cm}
\setlength{\belowdisplayskip}{0cm}
\text{H}_{\text{AS}}=\sqrt{\mathscr{L}_{\text{AS}}}\left( \sqrt{\frac{\iota _{\text{AS}}}{\iota _{\text{AS}}+1}}\overline{\text{H}}_{\text{AS}}+\sqrt{\frac{1}{\iota _{\text{AS}}+1}}\widetilde{\text{H}}_{\text{AS}} \right),
\end{align}
where $\iota _{\text{UA}}$ and  $\iota _{\text{AS}}$ represent the Rician factor for $\text{H}_{\text{UA}}$ and $\text{H}_{\text{AS}}$, respectively. $\text{diag}\left\{ \mathscr{L}_{\text{UA},1},\cdots ,\mathscr{L}_{\text{UA},K} \right\}  $ and  $\mathscr{L}_{\text{AS}}$ are the large-scale channel fading coefficients \cite{Huang2024IRS}. Besides, $\widetilde{\text{H}}_{\text{UA}}$ and $\widetilde{\text{H}}_{\text{AS}}$ are assumed to follow Rayleigh fading.
The LoS components in $\overline{\text{H}}_{\text{UA}}$ and $\overline{\text{H}}_{\text{AS}}$ are defined as
\begin{align}\label{h1LOS}
\setlength{\abovedisplayskip}{0cm}
\setlength{\belowdisplayskip}{0cm}
\overline{\text{H}}_{\text{UA}}=\alpha \left( \phi _{\text{UA}} \right) \alpha \left( \phi _{\text{AU}} \right) ,
\end{align}
\begin{align}\label{h2LOS}
\setlength{\abovedisplayskip}{0cm}
\setlength{\belowdisplayskip}{0cm}
\overline{\text{H}}_{\text{AS}}=\alpha \left( \phi _{\text{AS}} \right) \alpha \left( \phi _{\text{SA}} \right) ,
\end{align}
where $\alpha \left( \phi  \right)$ is the steering vector. $ \phi _{\text{UA}} $ and $\phi _{\text{AU}}$ are the direction-of-arrive (DoA) and direction-of-departure (DoD) angles, respectively. $ \phi _{\text{AS}} $ and $\phi _{\text{SA}}$ are the DoA and DoD angles, respectively. The steering vector $\alpha \left( \phi  \right)$ is defined as
\begin{align}\label{Steer}
\setlength{\abovedisplayskip}{0cm}
\setlength{\belowdisplayskip}{0cm}
\alpha \left( \phi  \right) =\left[ 1,e^{-j\frac{2\pi}{\lambda}d\sin\phi},\ldots,e^{-j\frac{2\pi}{\lambda}\left( N_I-1 \right) d\sin\phi} \right],
\end{align}
where $d={\lambda}/{2}$ is the distance between two adjacent AIRS reflecting elements.

\subsection{Computation Offloading Model} \label{computation}
We define the generated task of UEs as a 3-tuple: $( L,f_{l},\epsilon) $, where $L$ is the task workload generated at the beginning of each time interval. $f_{l}$ and $\epsilon$ denote the equivalent CPU frequency and the number of CPU cycles required to process a one-bit task, respectively.
In the MEC system, a partial computation offloading approach is adopted to process the computing tasks. At each short time interval $\tau$, UE$_k$ offloads partial computing tasks ${\chi}_{k}^{(\tau)}L_{k}^{(\tau)}$ to the AP for edge computing, where ${\chi}^{(\tau)}\in \left[ 0,1 \right]$ is the offloading ratio. At the same time, UE$_k$ is also capable of processing the rest of the tasks $(1-{\chi}_{k}^{(\tau)}) L_{k}^{(\tau)}$ locally with its own CPU.

Thus, the computational latency $T_{l,k}^{\left(\tau\right)}$ and energy consumption $E_{l,k}^{\left(\tau\right)}$ for the local computing of UE$_k$ at short time interval $\tau$ are determined by
\begin{equation}\label{Tloc}
\setlength{\abovedisplayskip}{0cm}
\setlength{\belowdisplayskip}{0cm}
T_{l,k}^{\left(\tau\right)}=\frac{\left(1-{\chi}_{k}^{(\tau)}\right) L_{k}^{(\tau)}\epsilon}{f_{l,k}},
\end{equation}
\begin{equation}\label{eloc}
\setlength{\abovedisplayskip}{0cm}
\setlength{\belowdisplayskip}{0cm}
E_{l,k}^{\left(\tau\right)}=\kappa\left(1-{\chi}_{k}^{(\tau)}\right) L_{k}^{(\tau)}\epsilon \left( f_{l,k}\right) ^2,
\end{equation}
where $\kappa $ is the capacitance coefficient determined by the CPU architecture and $f_{l,k}$ denotes the CPU frequency of UE$_k$.

Moreover, the latency $T_{o,k}^{\left(\tau\right)}$ and energy consumption $E_{o,k}^{\left(\tau\right)}$ in the offloading process of UE$_k$ at short time interval $\tau$ are determined by
\begin{equation}\label{Toff}
\setlength{\abovedisplayskip}{0cm}
\setlength{\belowdisplayskip}{0cm}
T_{o,k}^{\left(\tau\right)}=\frac{{\chi}_{k}^{(\tau)}L_{k}^{(\tau)}\epsilon}{f_{o}}+\frac{{\chi}_{k}^{(\tau)}L_{k}^{(\tau)}}{R_{k}^{(\tau)}},
\end{equation}
\begin{equation}\label{eoff}
\setlength{\abovedisplayskip}{0cm}
\setlength{\belowdisplayskip}{0cm}
E_{o,k}^{\left(\tau\right)}=\frac{p_{k}^{(\tau)}{\chi}_{k}^{(\tau)}L_{k}^{(\tau)}}{R_{k}^{(\tau)}},
\end{equation}
where $f_{o}$ is the CPU frequency of the MEC server.

\subsection{Problem Formulation}\label{problemformulation}
Since the local computing and the offloading process occur simultaneously, the total latency of UE$_k$ at the short time interval $\tau$ to complete the task is defined as $T_{k}^{\left(\tau\right)}=\max \{ T_{l,k}^{\left(\tau\right)}, T_{o,k}^{\left(\tau\right)} \}$. The total energy consumption is calculated as $E_{k}^{\left(\tau\right)}=E_{l,k}^{\left(\tau\right)}+E_{o,k}^{\left(\tau\right)} $.
As the DISCO jamming induces SINR degradation and task offloading reliability impairment, a penalty parameter $\varrho_{k}$ is introduced to penalize insecure offloading and determine whether the offloading process complies with strict latency requirements, which is defined as
\begin{equation}\label{failure}
\setlength{\abovedisplayskip}{0cm}
\setlength{\belowdisplayskip}{0cm}
\varrho _k=\left\{ \begin{array}{l}
	0,\ T_k\ge T_{\max}\\
	1,\ \text{otherwise}\\
\end{array} \right.,
\end{equation}
where $T_{\max}$ is the maximum delay restriction required by the computing task to fulfill the QoS of UE$_k$. The penalty parameter $\varrho_{k}$ penalizes latency violations caused by CSI corruption, and the energy efficiency term ensures the anti-jamming strategy is cost-effective. To comprehensively assess the task offloading performance under the DIRS-based FPJs, we define the secure offloading utility of UE$_k$ at a short time interval $\tau$ as
\begin{equation}\label{utility}
\setlength{\abovedisplayskip}{0cm}
\setlength{\belowdisplayskip}{0cm}
U_{k}^{\left(\tau\right)}=\varrho_{k}^{\left(\tau\right)} \frac{L_{k}^{\left(\tau\right)}}{E_{k}^{\left(\tau\right)}},
\end{equation}
where ${L_{k}^{\left(\tau\right)}}/{E_{k}^{\left(\tau\right)}}$ represents the energy efficiency for processing one bit of task.

To achieve the maximum system utility, we jointly optimize the computation offloading ratios $\boldsymbol{\chi} = [\chi_{1},\ldots,\chi_{k},\ldots,\chi_{K}]$, the phase shifts $\boldsymbol{\theta}_{A} = [\theta_{A,1},\ldots,\theta _{A, N_{A}}]$, and the location of AIRS  $\varUpsilon _A\left( x_A,y_A,z_A \right)$. We formulate the secure offloading utility maximization problem as follows:
\begin{equation}\label{formulation}
\begin{split}
\begin{array}{l}
\underset{\boldsymbol{\chi} , \boldsymbol{\theta }_{A} , {\varUpsilon _A}}{\max}\, \sum_{k=1}^{K}{U_k},\\
\mathrm{s.t.}\;\;({\rm{a}}):\;R_k\geq R_{\min}, k\in \left\{ 1,2,\ldots, K \right\},\\
\;\;\;\;\;\;\;({\rm{b}}):\;\chi_k\in \left[ 0,1 \right], k\in \left\{ 1,2,\ldots, K \right\},\\
\;\;\;\;\;\;\;({\rm{c}}):\;\varUpsilon _A\in \boldsymbol{\Upsilon },\\
\;\;\;\;\;\;\;({\rm{d}}):\;\theta _{A,1},...,\theta _{A,N_A}\in \left[ 0,2\pi \right),\\
\end{array}
\end{split}
\end{equation}
where $R_{\min}$ in (\ref{formulation}a) signifies the minimum offloading rate threshold at each UE \cite{Yao2024Anti-Jamming}. A higher offloading rate $R_k$ reduces latency and increases the probability that $\varrho_{k} = 1$, while the energy efficiency term ensures anti-jamming performance is achieved without excessive energy consumption, aligning with energy-efficient anti-jamming designs. (\ref{formulation}b)-(\ref{formulation}d) constrain the ranges for the parameters used above. $\boldsymbol{\Upsilon }$ denotes the discrete 3D location set.
The formulated problem in (\ref{formulation}) is high-dimensional and nonlinear, involving both discrete and continuous variables.
The complexity of the optimization problem arises from the inconsistent update frequency of the AIRS deployment and phase shift design. Besides, considering that the DIRS-jammed channels are time-varying, traditional non-convex optimization methods cannot effectively obtain the optimal AIRS deployment-aided secure offloading strategies, which rely on accurate mathematical models and perfect instantaneous CSI.
To this end, we propose a dual-agent DRL approach by integrating tools from the 2Ts framework and DRL algorithm in Section \ref{DDADSO}. Some important symbols are summarized in Table \ref{NotationTable}.
\begin{table}[!t]
    \setlength{\abovecaptionskip}{0pt}
    \setlength{\belowcaptionskip}{-6pt}
    \caption{List of Notations}
    \label{NotationTable}
    \centering
    \small
    \renewcommand{\arraystretch}{1.10}
    \begin{tabular}{>{\raggedright}p{2.2cm} p{5.8cm}}
        \toprule
        \textbf{Notation} & \textbf{Definition} \\
        \midrule
        \multicolumn{2}{l}{\textit{Main notations in the system model}} \\
        \midrule
        $E,\ E_l/E_o$         & Total energy consumption; energy consumption for local computing / offloading \\
        $\boldsymbol{h}$      & Equivalent effective channel \\
        $\mathrm{H}$          & Channel matrix \\
        $k$                   & Index of UE$_k$, $k\in\{1,2,\ldots,K\}$ \\
        $L$                   & Data size \\
        $N_A/N_D$             & Number of AIRS / DIRS reflecting elements \\
        $R_k$                 & Wireless offloading rate of UE$_k$ \\
        $T,\ T_l/T_o$         & Total computational latency; latency for local computing / offloading \\
        $t_i$                 & Long time interval of AIRS deployment, $i\in\{1,2,\ldots,I\}$ \\
        $\tau_j$              & Short time interval of AIRS-aided secure offloading,
                                $\tau_j\in[t_i,t_{i+1})$, $j\in\{1,2,\ldots,J\}$ \\
        $U_k$                 & Utility value of UE$_k$ \\
        $\varGamma_k$         & SINR for the offloading channel of UE$_k$ \\
        $\chi_k$              & Offloading ratio of UE$_k$ \\
        $\varUpsilon(x,y,z)$  & Three-dimensional Euclidean coordinate \\
        $\boldsymbol{\Theta}_A/\boldsymbol{\Theta}_D$ & AIRS / DIRS reflecting matrix \\
        $\boldsymbol{\theta}_A/\boldsymbol{\theta}_D$ & AIRS / DIRS reflecting phase shifts \\
        $\rho$                & Reflecting amplitude \\
        $\boldsymbol{\omega}$ & Beamforming matrix \\
        $\mathscr{L}$         & Large-scale channel fading coefficient \\
        \midrule
        \multicolumn{2}{l}{\textit{Main notations in the algorithm}} \\
        \midrule
        $\boldsymbol{a}$      & Action \\
        $\boldsymbol{s}$      & State \\
        $r$                   & Reward \\
        $R^{\text{out}}$      & Penalty item for constraint violation \\
        $\boldsymbol{z}$      & Low-dimensional latent variable \\
        $S_z$ & Dimension of latent variable \\
        $S_{in}$ & Dimension of VAE input state \\
        $n_{z}$,$n_{\mu}$,$n_{Q}$,$n_{\text{DQN}}$ & Hidden-layer width of the VAE, actor networks, critic networks, and DQN networks\\
        $J$                   & the number of short-timescale intervals within one long-timescale interval \\
        $\gamma_1,\ \gamma_2$ & Discount factors \\
        $\varpi$              & Parameters of neural networks \\
        \bottomrule
    \end{tabular}
\end{table}

\section{Dual-agent DRL-based AIRS deployment-Aided Secure Computation Offloading Scheme}\label{DDADSO}
In this section, we present the dual-agent DRL-based AIRS deployment-aided secure computation offloading (DDADSO) scheme against the DISCO jamming attacks. We begin by introducing the 2Ts optimization framework. Then, we outline and introduce the AIRS-aided computation offloading problem on long and short timescales, and formally define the related state, action, and reward spaces. Finally, the complexity analysis of the DDADSO algorithm is given.

\subsection{Two-Timescale Framework Decomposition}

\begin{figure}[!t]
\vspace{-0.7cm}
	\setlength{\abovecaptionskip}{0cm}
	\setlength{\belowcaptionskip}{-0.5cm}
	\begin{center}
		\includegraphics[height=0.9 in]{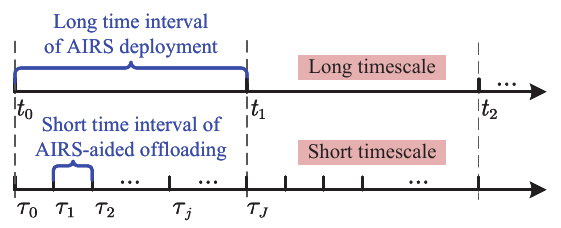} \\
		\caption{The specific example of 2Ts framework.}\label{2Tsframework}
	\end{center}
\end{figure}

The AIRS deployment strategy is primarily based on long-term environmental features, such as user distribution and the location of DIRS-based FPJs. Additionally, due to the high cost associated with physically redeploying AIRS, the AIRS deployment operates on a relatively long timescale. On the other hand, the joint optimization of the AIRS phase shifts and the computation offloading ratio is performed in real time based on the DIRS-jammed CSI and workloads. Therefore, the AIRS deployment and AIRS-aided computation offloading strategy are updated at different frequencies in the joint optimization problem. Hence, we adopt a 2Ts optimization framework and decompose the joint optimization problem into two subproblems: the long-timescale AIRS deployment problem and the short-timescale AIRS-aided secure computation offloading problem. An example of the 2Ts updating framework is shown in Fig. \ref{2Tsframework}.

\subsubsection{Long-Timescale AIRS Deployment Problem}
We optimize the AIRS deployment to provide secure LoS paths against the DISCO jamming attacks.
However, physically relocating or redeploying AIRS is costly and does require considerations such as energy consumption, system stability, and safety, which makes frequent adjustments impractical.  Furthermore, environmental changes related to deployment, such as user distribution and DIRS location, tend to evolve at a relatively gradual pace. Consequently, updating the AIRS deployment every few minutes or longer is sufficient to capture the macroscopic variations in the environment.
Consequently, this necessitates the implementation of a more robust optimization strategy.
As a result, we optimize the AIRS deployment on the long timescale, with update intervals possibly on the order of several quarters of an hour, to strengthen the guarantee of secure offloading against the DISCO jamming attacks. Specifically, we consider that the redeployment of AIRS is performed periodically at each long time interval, denoted by $t_i$, $i\in \left\{ 1,2,\ldots, I \right\}$.

\subsubsection{Short-Timescale AIRS-aided Secure Computation Offloading Problem}
Robust anti-jamming and efficient offloading performance require the joint optimization of offloading ratios and phase shifts in response to instantaneous CSI and current task requirements. Such frequent updates are feasible because that DIRS phase shifts remain constant within each update interval, and the legitimate UE-AIRS-AP channels are LoS-dominant Rician fading with high stability. Consequently, we update the AIRS-aided secure computation offloading strategy within each much shorter time interval $\tau_j$, $j\in \left\{ 1,2,\ldots, J \right\}$, typically on the order of milliseconds. This short-timescale optimization focuses on adjusting the offloading ratios and phase shifts to enhance secure offloading performance, complementing the long-term robustness achieved by the AIRS deployment strategy.

Accordingly, we decompose the AIRS deployment-aided secure computation offloading problem into sub-problems with different update frequencies, significantly improving secure offloading performance and reducing optimization complexity. Long-timescale AIRS deployment optimization achieves global control over DIRS-based FPJ-jammed environments, enabling short-timescale dynamic optimization to focus on rapid adaptation to CSI and task fluctuations, thereby enhancing system security stability and offloading efficiency.

\begin{figure}[!t]
	\setlength{\abovecaptionskip}{0cm}  
	\setlength{\belowcaptionskip}{-0.3cm}  
	\begin{center}
		\includegraphics[height=3.25 in]{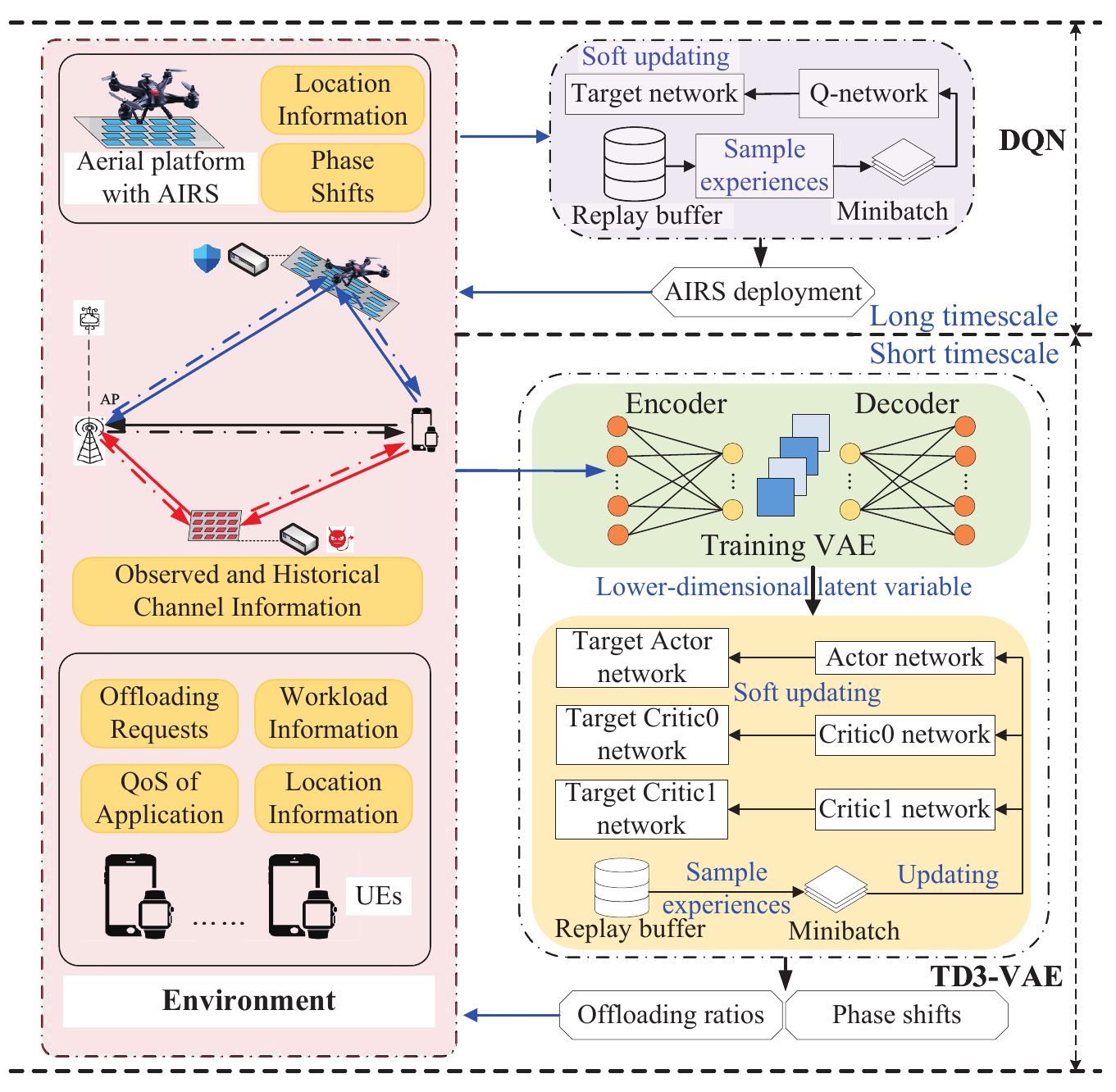}\
	\end{center}
	\caption{We propose a dual-agent DRL-based scheme that combines the DQN and TD3-VAE algorithms to determine the optimal AIRS deployment-aided secure computation offloading strategy.	}
	\label{2Ts-DRLframework}
\end{figure}

Given that the next states of the two sub-problems depend on their current states and actions, which aligns with the Markov property, they can be formulated as MDPs. Based on this, the optimal AIRS deployment-aided secure computation offloading strategy can be adaptively acquired through DRL techniques \cite{Liu2024Tcom, Yao2024Anti-Jamming}. The two-timescale decomposition aligns with the strengths of different DRL algorithms. Our proposed collaborative framework integrates VAE feature compression and DRL strategy optimization. VAE extracts low-dimensional core features to mitigate redundancy in the high-dimensional state space and enhance learning stability. This framework allows the dual-agent DRL scheme to use the TD3-VAE algorithm, significantly enhancing the efficiency and robustness of anti-jamming decisions. Specifically, the TD3-VAE-based DRL agent learns a mapping between observed offloading rate degradation and optimal phase shifts, avoiding explicit instantaneous CSI estimation of jamming channels. This reduces pilot overhead while ensuring adaptive tracking of jamming changes. As depicted in Fig. \ref{2Ts-DRLframework}, we propose the dual-agent DRL-based approach to acquire the optimal policy without requiring instantaneous perfect DIRS-related CSI.

\subsection{Long-Timescale AIRS Deployment}
On the long timescale, we optimize the AIRS deployment to strengthen the guarantee of secure offloading against the DISCO jamming attacks.
Then, we specify the state space, action space, and reward function of the MDP as follows:

\textbf{State Space}: In the AIRS-aided MEC framework, AIRS is appropriately deployed to assist the computation offloading in the presence of a DIRS by adjusting the location of the reflector. Thus, the system state $\boldsymbol{s}_{\text{long}}^{\left( t_i \right)}$ at each long time interval $t_i$ is modeled as
\begin{equation}\label{S}
\setlength{\abovedisplayskip}{0cm}
\setlength{\belowdisplayskip}{0cm}
\boldsymbol{s}_{\text{long}}^{\left( t_i \right)}=\left[ \varUpsilon _{A}^{\left( t_{i-1} \right)}, {\text{H}_{\text{DT}}}^{\left( t_{i-1} \right)}\right],
\end{equation}
where $\varUpsilon _{A}^{\left( t_{i-1} \right)}$ is the AIRS's location at last long time interval $t_{i-1}$. $\text{H}_{\text{DT}}=[ \boldsymbol{h}_{\text{DT},1},\ldots,\boldsymbol{h}_{\text{DT},K}] $, $\forall k\in \left\{ 1,2,\ldots,K \right\}$ denotes the estimated CSI of UEs based on the $ \varUpsilon _{A}^{\left( t_{i-1} \right)}$.
It should be noted that the channel coefficients and the SINR $\boldsymbol{\varGamma}^{\left( t_{i} \right)}=[ {\varGamma}_{1}^{\left( t_{i} \right)},\ldots,{\varGamma}_{K}^{\left( t_{i} \right)}] $, $\forall k\in \left\{ 1,2,\ldots,K \right\}$ are significantly influenced by the locations of AIRS. We utilize the historical location information and CSIs as part of the state to learn the policy \cite{Liu2024Tcom, Yao2024Anti-Jamming}.

\textbf{Action Space}: The AIRS deployment has a significant impact on the overall network topology, including AIRS-aided offloading channels and anti-jamming capability against the DISCO jamming attacks. The long-timescale action space $\boldsymbol{a}_{\text{long}}^{\left( t_i \right)}$ can be defined as follows:
\begin{equation}
\label{Aslow}
  \boldsymbol{a}_{\text{long}}^{\left( t_i \right)}=\varUpsilon _{A}^{\left( t_{i} \right)}.
\end{equation}

\textbf{Reward}: The reward function assesses the AIRS deployment strategy when the DQN-based agent executes an action decision according to the present state. To maximize the energy efficiency while ensuring that all UEs meet the delay threshold and the minimum offloading rate requirements, we denote the sum of secure offloading utility as the reward. This design enables the deployment policy to improve the anti-jamming capability. Then, the reward $r_{\text{long}}^{\left( t_i \right)}$ at each long time interval $t_i$ is modeled as
\begin{equation}\label{Rhigh1}
\setlength{\abovedisplayskip}{0cm}
\setlength{\belowdisplayskip}{0cm}
r_{\text{long}}^{\left( t_i \right)}=\sum\limits_{j=1}^J{\sum\limits_{k=1}^K{r_{k}^{\left( \tau _j \right)}}},
\end{equation}
\begin{equation}\label{Rhigh2}
\setlength{\abovedisplayskip}{0cm}
\setlength{\belowdisplayskip}{0cm}
r_{k}^{\left( \tau _j \right)}=\lambda _1U_{k}^{\left( \tau _j \right)}+\lambda _2R_{k}^{\text{out}},
\end{equation}
where $\tau _j\in \left[ t_i,t_{i+1} \right) $ denotes short time intervals in the long time interval $t_i$. Besides, $\lambda _1$ and $\lambda _2$ are the weighting parameters. $R_{k}^{\text{out}}$ denotes a penalty item and is designed to stabilize training \cite{Yao2024Anti-Jamming}, which is given by
\begin{equation}\label{Rout}
\setlength{\abovedisplayskip}{0cm}
\setlength{\belowdisplayskip}{0cm}
\begin{aligned}
R_{k}^{\text{out}}=\left\{ \begin{array}{l}
	0,\ \ \ \ R_{k}^{\left( \tau _j \right)}\geq R_{\min}\\
	-R_{k}^{\left( \tau _j \right)},\ 0<R_{k}^{\left( \tau _j \right)}<R_{\min}\\
	-1,\ \ \ R_{k}^{\left( \tau _j \right)}=0\\
\end{array} \right..
\end{aligned}
\end{equation}

The DQN-based agent utilizes the DQN algorithm to design the AIRS deployment on the long timescale, where the location is selected in a discrete and finite set. In addition, the DQN-based agent uses experience replay and an independent target network to address algorithmic instability. Specifically, the AIRS deployment experiences $\left\{ \boldsymbol{s}_{\text{long}}^{\left( t_i \right)},\boldsymbol{a}_{\text{long}}^{\left( t_i \right)}, {r}_{\text{long}}^{\left( t_i \right)},\boldsymbol{s}_{\text{long}}^{\left( t_{i+1} \right)} \right\}$ are stored in the replay buffer, allowing data sampling to train the DQN network.
The DQN-based agent employs an evaluation neural network $Q\left( \boldsymbol{s, a}\mid \varpi \right)$ with the parameter $ \varpi$ to assess the Q-value and uses a target neural network $Q\left( \boldsymbol{s, a}\mid \varpi '\right)$ with the parameter $ \varpi'$ to update the evaluation network \cite{Yu2021When}. The DQN loss function is defined as follows:
\begin{equation}\label{DQN}
	\begin{aligned}
	y_{\text{long}}^{\left( t_i \right)} &=r_{\text{long}}^{\left( t_i \right)}+\gamma_1\max_{\boldsymbol{a}}Q\left( \boldsymbol{s}_{\text{long}}^{\left( t_{i+1} \right)},\boldsymbol{a;}\varpi ' \right),\\
	\mathscr{Z}\left( \varpi \right) &=\mathbb{E}\left[ \left( y_{\text{long}}^{\left( t_i \right)}-Q\left( \boldsymbol{s}_{\text{long}}^{\left( t_i \right)},\boldsymbol{a;}\varpi \right) \right) ^2 \right],\\
\end{aligned}
\end{equation}
where $\gamma_1$ is the discount factor of the DQN. The parameters of the target network $\varpi '$ are periodically updated with the parameters from the parameter of evaluation network $\varpi$ \cite{Yu2021When}.

\subsection{Short-Timescale AIRS-Aided Secure Computation Offloading}
\subsubsection{MDP model}
The AP, as a TD3-based agent, dynamically selects the offloading ratios of UEs while designing the AIRS phase shifts on the short timescale.
The fundamental components are outlined as follows:

\textbf{State}: UEs send the computation offloading requests to the AP, encompassing workload information, channel information, and QoS information for specific applications.
Then, the AP receives the uploaded information from UEs and integrates the observed and historical channel information \cite{Yao2024Anti-Jamming}. Thus, the system state $\boldsymbol{s}_{\text{short}}^{\left( \tau_j \right)}$ at each short time interval $\tau_{j}$ is modeled as
\begin{equation}\label{S}
\setlength{\abovedisplayskip}{0cm}
\setlength{\belowdisplayskip}{0cm}
\boldsymbol{s}_{\text{short}}^{\left( \tau _j \right)}=\left[ \boldsymbol{L}^{\left( \tau _j \right)},\text{H}_{\text{AS}}^{\left( \tau _j \right)},\text{H}_{\text{UA}}^{\left( \tau _j \right)},\boldsymbol{R}^{\left( \tau _{j-1} \right)} \right],
\end{equation}
where $\boldsymbol{L}^{\left( \tau _j \right)}=[ L_{1}^{\left( \tau _j \right)},\ldots,L_{k}^{\left( \tau _j \right)},\ldots,L_{K}^{\left( \tau _j \right)} ] $ denotes the workloads of UEs. $\text{H}_{\text{AS}}^{\left( \tau _j \right)}$ and $\text{H}_{\text{UA}}^{\left( \tau _j \right)}$ represent the current channel state. $\boldsymbol{R}^{\left( \tau _{j-1} \right)}$ is the previous offloading rate.

\textbf{Action}:
The AIRS-aided secure computation offloading strategy $\boldsymbol{a}_{\text{short}}^{ ( \tau_j  )}$ includes computation offloading ratios and the AIRS reflecting coefficient matrix  at each short time interval $\tau_{j}$. Hence, the action of the TD3-based agent is defined as
\begin{equation}\label{A}
\setlength{\abovedisplayskip}{0cm}
\setlength{\belowdisplayskip}{0cm}
\boldsymbol{a}_{\text{short}}^{\left( \tau_j \right)}=\left[ \boldsymbol{\chi }^{\left( \tau_j \right)},\boldsymbol{\theta }_{A}^{\left( \tau_j \right)} \right] .
\end{equation}

\textbf{Reward}:
The utility value of the offloading performance increases as the reward aligns proportionally with the specific objective delineated in (\ref{formulation}). We directly integrate the latency thresholds and the minimum offloading rate requirement embedded in the utility function. Hence, the reward function for the AIRS-aided secure computation offloading strategy is presented as follows:
\begin{equation}\label{Rlow}
\setlength{\abovedisplayskip}{0cm}
\setlength{\belowdisplayskip}{0cm}
{r}_{\text{short}}^{\left( \tau_j \right)}=\sum\limits_{k=1}^K{r_{k}^{\left( \tau _j \right)}}.
\end{equation}

The TD3-based agent dynamically designs the AIRS phase shifts and the offloading ratios of UEs to adapt to instantaneous CSI and current task demands. Furthermore, the TD3-VAE-based AIRS-aided secure computation offloading approach is introduced to address the high-dimensional and continuous-action problem. Specifically, the VAE model is designed to help the TD3-based agent obtain a low-dimensional representation of its environment \cite{Larsen2024VAE, Sun2025GAI}. As a result, we can get a higher probability that the TD3 algorithm will find a good AIRS-aided secure computation offloading policy.

\subsubsection{VAE architecture}
Similar to the design in \cite{Sun2025GAI}, we implement fully connected layers as feature extractors. As shown in Table \ref{parameterVAE}, the encoder in VAE adopts the fully connected layer architecture, and the decoder uses its transposed versions.
\begin{table}[!t]
	\centering
	\caption{Parameters of The VAE.}
	\label{parameterVAE}
    \small
	\renewcommand{\arraystretch}{1.1} 
	\begin{tabular*}{0.48\textwidth}{p{0.24\textwidth} p{0.20\textwidth}} 
		\toprule
		Parameter & Configurations \\
		\midrule
 		No. layers & 2 \\
        Dimension of latent variable & $4$ \\
        Number of hidden layer nodes & 16(Layer 1) 8(Layer 2) \\
        Activation function & LeakyReLU\\
		\bottomrule
	\end{tabular*}
\end{table}
The above model is implemented and trained on approximately 80,000 samples with a learning rate of 0.0001 and a batch size of 64 for 20 epochs using the Adam optimizer.

\subsubsection{TD3 algorithm for the short-timescale AIRS-aided secure computation offloading}
Based on the selected AIRS-aided secure computation offloading strategy, the TD3-based agent receives an instant reward ${r}_{\text{short}}^{ ( \tau_j  )}( \boldsymbol{s}_{\text{short}}^{ ( \tau_j  )}, \boldsymbol{a}_{\text{short}}^{ ( \tau_j  )} )$ and transits the system state from $\boldsymbol{s}_{\text{short}}^{ ( \tau_j  )}$ to $\boldsymbol{s}_{\text{short}}^{ ( \tau_{j+1} )}$. The TD3 algorithm employs six deep neural networks: one Actor network, one target Actor network, two Critic networks, and two target Critic networks. In particular, $\varpi ^{\mu}$ and $\varpi^{Q} _i$ are the parameters for the Actor network and Critic networks, while $\varpi ^{\mu'} $ and $\varpi^{Q'} _{{i}}$ are the parameters for the target Actor network and target Critic networks, $i=\,\,1,2$ \cite{Zhang2025IoT}. Specifically, the Actor network $A_{\varpi ^{\mu}}$ generates the AIRS-aided secure computation offloading strategy \(\boldsymbol{a}_{\text{short}}^{(\tau_j)}\) based on the current state $\boldsymbol{s}_{\text{short}}^{(\tau_j)}$, while the target Actor network $A_{\varpi ^{\mu'} }$ outputs the target action $\widetilde{\boldsymbol{a}}_{\text{short}}^{(\tau_j)}$ based on the next state $\boldsymbol{s}_{\text{short}}^{(\tau_{j+1})}$. The Critic network $Q_{\varpi^{Q} _i}$ calculates the Q value $Q_{\varpi^{Q} _i}( \boldsymbol{s}_{\text{short}}^{( \tau_j )},\boldsymbol{a}_{\text{short}}^{\left( \tau_j \right)} )$ based on the state $\boldsymbol{s}_{\text{short}}^{(\tau_j)}$ and action \(\boldsymbol{a}_{\text{short}}^{(\tau_j)}\), while the target Critic network $Q_{\varpi^{Q'} _i}$ calculates the Q value $Q_{\varpi^{Q'} _i}( \boldsymbol{s}_{\text{short}}^{(\tau_{j+1})}, \widetilde{\boldsymbol{a}}_{\text{short}}^{( \tau_j )} ) $ based on the state $\boldsymbol{s}_{\text{short}}^{(\tau_{j+1})}$ and action $\widetilde{\boldsymbol{a}}_{\text{short}}^{(\tau_j)}$. The TD3-based agent stores the AIRS-aided secure computation offloading experiences $\left\{ \boldsymbol{s}_{\text{short}}^{\left( \tau_j \right)}, \boldsymbol{a}_{\text{short}}^{\left( \tau_j \right)}, {r}_{\text{short}}^{\left( \tau_j \right)},\boldsymbol{s}_{\text{short}}^{\left( \tau_{j+1} \right) }  \right\} $ in the replay buffer for network parameter update.

When updating the critic network parameters, the minimum Q value from the two critic target networks is selected as the target value $y^{\left( k \right)}$, that is:
\begin{equation}\label{updateTD31}
\setlength{\abovedisplayskip}{0cm}
\setlength{\belowdisplayskip}{0cm}
\begin{aligned}
\left\{ \begin{array}{l}
	\widetilde{\boldsymbol{a}}_{\text{short}}^{\left(\tau_j \right)}=A_{\varpi ^{\mu'}}\left( \boldsymbol{s}_{\text{short}}^{\left( \tau_{j+1} \right)} \right) +\varepsilon\\
	y^{\left(\tau_j\right)}=r_{\text{short}}^{\left(\tau_j \right)}+\gamma_2 \underset{i=1,2}{\min}Q_{\varpi^{Q'} _i}\left( \boldsymbol{s}_{\text{short}}^{\left( \tau_{j+1} \right)},\widetilde{\boldsymbol{a}}_{\text{short}}^{\left(\tau_j \right)} \right)\\
	\Delta \varpi^{Q} _i=\nabla _{\varpi^{Q} _i}\left( y^{\left(\tau_j  \right)}-Q_{\varpi^{Q} _i}\left( \boldsymbol{s}_{\text{short}}^{\left(\tau_j\right)},\boldsymbol{a}_{\text{short}}^{\left(\tau_j\right)} \right) \right) ^2\\
\end{array} \right.,
\end{aligned}
\end{equation}
where $\varepsilon$ is the exploration noise, which is independently sampled from a truncated normal distribution. $\gamma_2$ represents the discount factor of the TD3.
Furthermore, the parameters of the actor network and the four target networks are updated using the following formula:
\begin{equation}\label{updateTD32}
\setlength{\abovedisplayskip}{0cm}
\setlength{\belowdisplayskip}{0cm}
\begin{aligned}
\left\{ \begin{array}{l}
	\Delta \varpi ^{\mu} =\nabla _{\varpi ^{\mu}}Q_{\varpi^{Q} _i}\left( \boldsymbol{s}^{\left(\tau_j\right)},A_{\varpi ^{\mu}}\left( \boldsymbol{s}^{\left(\tau_j\right)} \right) \right)\\
	\varpi^{Q'} _i\gets \psi \varpi^{Q} _i+\left( 1-\psi  \right) \varpi^{Q'} _i\\
	\varpi ^{\mu'}\gets \psi  \varpi ^{\mu}+\left( 1-\psi  \right) \varpi ^{\mu'}\\
\end{array} \right.,
\end{aligned}
\end{equation}
where $\psi $ is the updating factor.

\begin{algorithm}[!t]

    \caption{TD3-VAE-based AIRS aided Secure Computation Offloading Scheme}
    	\setlength{\abovecaptionskip}{0cm}  
	\setlength{\belowcaptionskip}{0cm}  
    \label{alg1}
    \begin{algorithmic}[1]
            \STATE Initialize VAE encoder $q_{\varpi^q}$ and decoder $p_{\varpi^p}$;
            \STATE Initialize $\boldsymbol{s}_{\text{short}}^{(\tau_j)}$, $\varpi^\mu$, and $\varpi_i^Q$;
            \STATE \textbf{if} $i\leq N_{\boldsymbol{z}}$ \textbf{then}
            \STATE \quad Sample the latent variable $\boldsymbol{z}_i$ ;
            \STATE \quad Pre-train VAE by optimizing $(\varpi^q,\varpi^p)$ using (\ref{33}) ;
            \FOR{$\tau_j=\tau_1,\tau_2,\ldots,\tau_J$}
                \STATE  Obtain $\boldsymbol{s}_{\text{short}}^{(\tau_j)}$ and current latent representation  $\boldsymbol{z}^{\left( \tau _j \right)}$;
                \STATE  Select the AIRS-aided secure computation offloading strategy $\boldsymbol{a}_{\text{short}}^{\left( \tau_j \right)}$ according to (\ref{34});
                \STATE  Obtain ${r}_{\text{short}}^{\left( \tau_j \right)}$ and observe the next state $\boldsymbol{s}_{\text{short}}^{\left(  \tau_{j+1}\right)}$;
                \STATE  Store the AIRS-aided secure computation offloading experiences $\{ \boldsymbol{s}_{\text{short}}^{\left( \tau_j \right)}, \boldsymbol{a}_{\text{short}}^{\left( \tau_j \right)}, {r}_{\text{short}}^{\left( \tau_j \right)},\boldsymbol{s}_{\text{short}}^{\left( \tau_{j+1} \right) }  \} $;
                \STATE  Extract experiences from replay buffer to update the deep neural networks of the TD3 algorithm according to (\ref{updateTD31}) and (\ref{updateTD32});
                \STATE  Update the short-timescale state $\boldsymbol{s}_{\text{short}}^{(\tau_j)} \leftarrow \boldsymbol{s}_{\text{short}}^{\left(  \tau_{j+1}\right)}$;
            \ENDFOR
    \end{algorithmic}
\end{algorithm}

\subsubsection{TD3-VAE-based AIRS aided Secure Computation Offloading Algorithm}
During each training iteration, we input the high-dimensional state data $\boldsymbol{\boldsymbol{s}}$ into the encoder of the VAE to extract a lower-dimensional latent variable $\boldsymbol{z}$. Specifically, the encoder network maps input data to a variational latent distribution $ q_{\varpi^{q}}\left( \boldsymbol{z}\left| \boldsymbol{s}\right. \right)$, which comprises a probabilistic representation of the underlying structure in the data. During training, the VAE aims to minimize the negative evidence lower bound (ELBO), which consists of a reconstruction term and a KL regularization term. Specifically, the encoder maps the input state to the variational posterior $q_{\varpi^q}(\boldsymbol{z}|\boldsymbol{s})$, while the decoder reconstructs the input through $p_{\varpi^p}(\boldsymbol{s}|\boldsymbol{z})$. The loss function is given by
\begin{equation}\label{33}
    \begin{aligned}
        Z(\varpi^q,\varpi^p)
        &= \sum_{i=1}^{N_z}
        \left(
        -\mathbb{E}_{\boldsymbol{z}_i\sim q_{\varpi^q}(\boldsymbol{z}_i|\boldsymbol{s}_i)}
        \left[\log p_{\varpi^p}(\boldsymbol{s}_i|\boldsymbol{z}_i)\right]\right.\\
        &\left.\quad+ D_{\mathrm{KL}}
        \left(
        q_{\varpi^q}(\boldsymbol{z}_i|\boldsymbol{s}_i)\,\|\,p(\boldsymbol{z}_i)
        \right)
        \right).
    \end{aligned}
\end{equation}
The first term is the reconstruction term, while the second term regularizes the variational posterior toward the predefined prior distribution.
where $N_{\boldsymbol{z}}$ represents the training number of the latent variable. The previous term ensures appropriate latent variable generation, while the latter KL divergence term makes the latent representation distribution match the prior distribution. During the policy optimization phase, the encoder generates the latent representation $\boldsymbol{z}$ from the input state $\boldsymbol{s}_{short}$. Then, the actor network $A_{\varpi^\mu}$ selects the AIRS-aided secure computation offloading strategy based on $\boldsymbol{z}$. This process is denoted by
\begin{equation}\label{34}
	\begin{aligned}
	\boldsymbol{z}&\sim q_{\varpi ^q}\left( \cdot \left| \boldsymbol{s}_{\text{short}} \right. \right),\\
	\boldsymbol{a}_{\text{short}}&\sim A_{\varpi ^{\mu}}\left( \boldsymbol{z} \right) +\varepsilon.\\
\end{aligned}
\end{equation}
The TD3-VAE algorithm is presented in the Algorithm \ref{alg1}.

In summary, as shown in Algorithm \ref{alg2}, the decision of the DQN-based agent plays an important role in guiding the policy selection of the TD3-based agent. In particular, the reward function of the long-timescale agent is designed as the sum of all short-timescale rewards within the entire long interval, ensuring that the long-timescale deployment optimization is guided by stable and representative system performance. Additionally, the short-timescale agent adopts a VAE encoder to extract low-dimensional latent features from high-dimensional and redundant states. Thus, the VAE-aided dual-agent DRL framework effectively alleviates non-stationarity in interactive dual-agent learning.

\vspace{-0.3cm}    
\begin{algorithm}[!t]

    \caption{Dual-agent DRL-based AIRS Deployment-aided Secure Computation Offloading Scheme (DDADSO)}
    	\setlength{\abovecaptionskip}{0cm}  
	\setlength{\belowcaptionskip}{0cm}  
    \label{alg2}
    \begin{algorithmic}[1]
            \STATE Initialize   $\eta $, $\varpi ^{\mu}$, $\varpi ^Q$, and the exploration noise $\mathcal{N} $;
            \STATE Initialize  $\varpi $ and $ \varUpsilon _{A}^{\left( t_{0} \right)}$ ;
            \FOR{episode = 1, 2, 3, \ldots}
            \STATE  Obtain the initial state $\boldsymbol{s}_{\text{long}}^{\left( t_1 \right)}= [ \varUpsilon _{A}^{\left( t_{0} \right)}, {\text{H}_{DT}}^{\left( t_{1} \right)} ]$;
            \FOR{ $t_i$ = $t_1$, $t_2$,  \ldots , $t_I$ }
                 \STATE Select the AIRS deployment strategy $ \boldsymbol{a}_{\text{long}}^{\left( t_i \right)}=\varUpsilon _{A}^{\left( t_{i} \right)}$;	
                 \STATE Obtain the optimal AIRS-aided secure computation offloading policy and the secure offloading utility of UEs by Algorithm \ref{alg1};
                 \STATE  Evaluate the average secure offloading utility and obtain the long-timescale reward value $r_{\text{long}}^{\left( t_i \right)}$;
                 \STATE  Observe the next state $\boldsymbol{s}_{\text{long}}^{\left( t_{i+1} \right)}$;
                 \STATE  Store $\left\{ \boldsymbol{s}_{\text{long}}^{\left( t_i \right)},\boldsymbol{a}_{\text{long}}^{\left( t_i \right)}, {r}_{\text{long}}^{\left( t_i \right)},\boldsymbol{s}_{\text{long}}^{\left( t_{i+1} \right)} \right\}$ as AIRS deployment experiences into the replay buffer;
                 \STATE  Extract experiences from the replay buffer to update the network parameters $\varpi $ and $\varpi '$;
                 \STATE  Update the long-timescale state $\boldsymbol{s}_{\text{long}}^{(t_i)} \leftarrow \boldsymbol{s}_{\text{long}}^{\left(t_{i+1}\right)}$;
            \ENDFOR
        \ENDFOR
    \end{algorithmic}
\end{algorithm}

\subsection{Complexity Analysis}
To further validate the feasibility of the proposed DDADSO algorithm, we conduct a detailed computational complexity analysis in the following. The notations used in this section have been summarized in Table \ref{NotationTable}.

\emph{1) Complexity of the TD3-VAE algorithm for short-timescale optimization:}
According to Table \ref{parameterVAE}, the VAE has two fully connected layers, with $n_{z}$ as its hidden-layer width. Thus, its encoder-decoder complexity is given by $\mathcal{O}\!\left(S_{in}n_z+n_z^2+n_zS_z\right)$, where $S_{in}$ and $S_z$ are the dimensions of the VAE input state and latent variable, respectively. For the TD3 algorithm, with fixed latent-state and action dimensions, the dominant complexity comes from the hidden-layer multiplications of the actor and critic networks \cite{Zhang2025IoT}. Hence, the computational complexity of the TD3 algorithm can be approximated as $\mathcal{O}\!\left(n_{\mu}^2+n_Q^2\right)$, where $n_{\mu}$ and $n_{Q}$ are the hidden-layer widths of the actor networks and critic networks, respectively. Therefore, the complexity of the TD3-VAE algorithm for short-timescale optimization is
\begin{equation}
\mathcal{O}\!\left(S_{in}n_z+n_zS_z+n_z^2+n_{\mu}^2+n_Q^2\right).
\label{eq:complexity_short}
\end{equation}

\emph{2) Complexity of the DQN algorithm for long-timescale optimization:}
The DQN evaluation network employs two fully connected hidden layers, with $n_{DQN}$ as its hidden-layer width. According to \cite{wang2022energy}, the complexity of the DQN algorithm is approximated as
\begin{equation}
\mathcal{O}\!\left(n_{DQN}^2\right).
\label{eq:complexity_long}
\end{equation}

Since one long-timescale interval contains $J$ short-timescale updates, according to \cite{Chen2024}, the overall computational complexity of DDADSO per long-timescale interval is
\begin{equation}
\mathcal{O}\!\left(J\left(S_{in}n_z+n_zS_z+n_z^2+n_{\mu}^2+n_Q^2\right)+n_{DQN}^2\right).
\label{eq:complexity_total}
\end{equation}
Notably, there is an increase in computational complexity compared to the single-timescale baselines. Despite this, we leverage the core advantage of the MEC architecture and deploy the MEC server as the execution agent of our algorithm. Hence, DDADSO does not impose heavy computational burden on UEs, and it reduces redundancy in the high-dimensional state space via the VAE module.

\section{Performance Evaluation}\label{SimulationResults}

\subsection{Simulation Settings}
We consider a three-dimensional Euclidean coordinate scenario to describe the location of UE$_k$ $\varUpsilon _{k} \left( x_{k},y_{k},z_{k} \right)$ and the AIRS center $\varUpsilon _A\left( x_A,y_A,z_A \right)$, which are measured in meters \cite{hu2021reconfigurable}. For simplicity, we consider a multi-user MEC system where $3$ UEs are assisted by an AIRS with 128 reflecting elements to offload partial computing tasks to the 12-antenna AP, and the DIRS with 1024 reflecting elements cause DISCO jamming attacks, i.e., $K=3$, $N_A=128$, $M=12$, and $N_D=1024$ \cite{ AIRS2022Niu}. Other parameters for UEs are detailed in Table \ref{parameter}. A one-bit DIRS is considered as the DIRS-based FPJ and the simulated amplitude and phase are determined by the specific design of the DIRS element, where the phase shifts and the corresponding amplitudes randomly chosen from $\left\{ \frac{\pi}{6},\frac{2\pi}{3} \right\}$ and $\left\{0.5,0.8\right\}$\cite{HuangHuan2024_5}. We consider that the altitude range of the AIRS service is $z_A\in \left[ 100,400 \right]$ m \cite{Cheng2022RotationLocation, Shang2022AIRS}. We then set a square area with a width of 500 meters as the operational space for the AIRS, and the discrete intervals are set to 50 m \cite{Wang2024}.
For each channel, the Rician factors $\iota _{DS}$, $\iota _{UA}$, and $\iota _{AS}$ are set to 10. The propagation parameters of the offloading channels are described as follows: $\mathscr{L}_{DS}$, $\mathscr{L}_{AS}$ = $35.6+20\log _{10}\left( d_i \right)$ and $\mathscr{L}_{US,k} $, $\mathscr{L}_{UD,k}$, $\mathscr{L}_{UA,k}$= $32.6+30\log _{10}\left( d_i \right)$, where $d_i\in \left\{ d_{DS },d_{AS },d_{US,k},d_{UD,k},d_{UA,k} \right\} $ \cite{HuangHuan2024_5}.

\begin{table}[!t]
	\centering
	\caption{Parameter Settings for Simulation.}
	\label{parameter}
	\renewcommand{\arraystretch}{1.1} 
    \small
	\begin{tabular*}{0.48\textwidth}{p{0.28\textwidth} p{0.20\textwidth}} 
		\toprule
		Parameter & Value \\
		\midrule
 		Task size $L$& 20-40 Kb \\
        Transmit power $p$&$2$ dB\\
        Average bandwidth of all UEs $B$&$2$ MHz\\
        CPU frequency of all UEs ${{f}_l}$&$4\times 10^9$ cycles/s\\
        CPU frequency of MEC server ${{f}_o}$&$8\times 10^{10}$ cycles/s\\
        Number of CPU cycles per bit $\epsilon$& 800 cycles/bit \\
        Consumed energy per CPU cycle $\kappa $&$1\times 10^{\left( -28 \right)} $ J/cycle\\
        Background noise power $\sigma ^{2}$ & -80 dBm \\
		\bottomrule
	\end{tabular*}
\end{table}

We conduct extensive experimentation and hand-tuning to identify the hyperparameters for the Dual-agent DRL-based algorithm in the DDADSO scheme.
Specifically, the learning rates of actor and critic networks in the TD3-based agent are 0.0001 and 0.00002, respectively. The size of the replay buffer and minibatch of TD3 are set to 10000 and 64 according to the algorithm training requirements in the DDADSO scheme. Besides, we set the learning rate and discount factor of the DQN algorithm as $ 0.0001$ and $\gamma=0.9$, respectively.
Then, we conduct simulations of the Dual-agent DRL-based algorithm utilizing Python 3.9 and PyTorch 2.6.0.

Furthermore, to underscore the performance advantages of the DDADSO scheme, we employ three benchmarks for comparative analysis. The detailed setup of each benchmark is described below.

\textbf{DDADSO (DDPG-based scheme)}: We propose a dual-agent DRL algorithm that combines the DQN and DDPG algorithms. Specifically, the DDPG-based agent jointly optimizes the computation offloading and phase shifts of AIRS.

\textbf{DICO W/O AD} \cite{ wang2022resource}: The DRL-based IRS-aided computation offloading scheme in \cite{ wang2022resource} is modified and applied in the AIRS-aided MEC system without AIRS deployment, named DICO W/O AD, as a benchmark. Specifically, the DICO W/O AD scheme jointly optimizes the computation offloading and phase shifts of AIRS, without considering AIRS deployment.

\textbf{DADCO} \cite{Shang2022AIRS}: This AIRS-assisted MEC scheme proposed in \cite{Shang2022AIRS} jointly optimizes the AIRS trajectory and phase shifts to enhance the offloading performance. On this basis, we modify this scheme and optimize the AIRS deployment, phase shifts, and offloading ratios, without distinguishing between fast and slow timescales.
Then, we name this scheme DADCO and apply it in the AIRS-aided MEC system against DISCO jamming attacks.

\textbf{Fixed AIRS}: The AIRS phase shifts and deployment are fixed at the pre-defined design, and the offloading ratios are optimized with the TD3 algorithm.

\textbf{Random AIRS}: The AIRS phase shifts and deployment are randomly generated, and the offloading ratios are optimized with the TD3 algorithm.

\subsection{Impact of DISCO Jamming Attacks on Computation Offloading}
First, we studied the influence of transmission power control on communication performance under different jamming attacks to prove its ineffectiveness against DISCO jamming. Then, we examine the effectiveness of AIRS-aided offloading in mitigating DISCO jamming, which supports the inclusion of AIRS phase shift optimization.

\begin{figure}[!t]
\vspace{-0.3cm}
	\setlength{\abovecaptionskip}{0cm}
	\setlength{\belowcaptionskip}{-0.5cm}
	\begin{center}
		\includegraphics[width= 2.5 in]{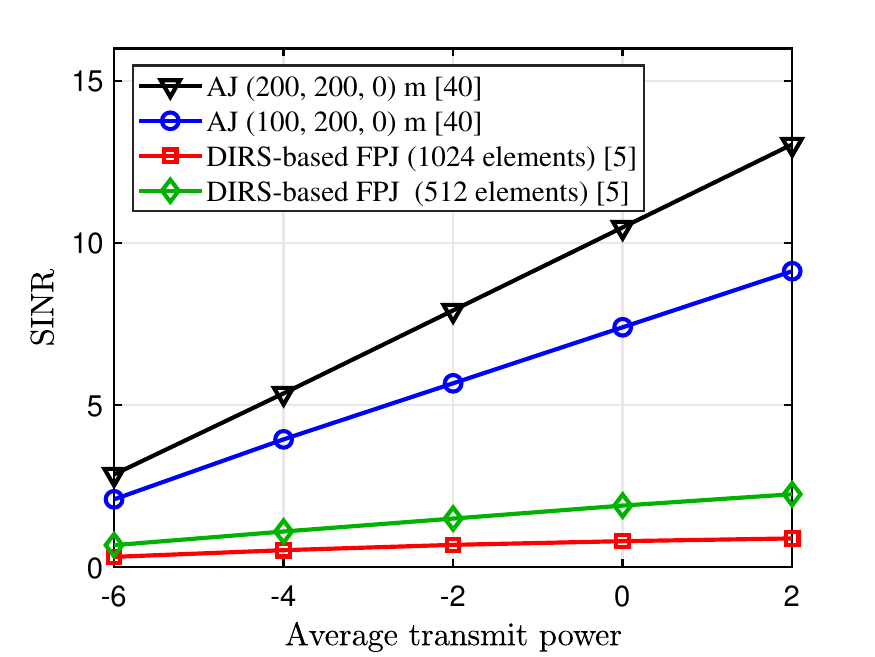} \\
		\caption{SINR versus power control under different jamming attacks.}\label{Fig4}
	\end{center}
\end{figure}
\subsubsection{SINR versus power control under different jamming attacks}
Fig. \ref{Fig4} illustrates the impact of transmit power control on SINR under different jamming attacks, including the active jammer (AJ) \cite{Liu2024Jamming} and the DIRS-based FPJ \cite{HuangFPJ2024}. We vary the transmit power $P$ from 4 W to 20 W, considering two AJs at different locations and two DIRS-based FPJs with 512 and 1024 reflecting elements. Increasing $P$ significantly improves SINR under AJ attacks, confirming the effectiveness of power control against active jamming. In contrast, for DISCO jamming, $P$ increments yield limited SINR gains, especially with more reflecting elements. For example, SINR increases by 4.56 times under an AJ at (200, 200, 0) m, but only 2.79 times for the 1024-element DIRS-based FPJ. This highlights the fundamental difference between AJ and DIRS-based FPJ. DIRS disrupts CSI acquisition of users and renders traditional power control strategies largely ineffective.

\begin{figure}[!t]
	\setlength{\abovecaptionskip}{0cm}
	\setlength{\belowcaptionskip}{-0.5cm}
	\begin{center}
		\includegraphics[width= 2.5 in]{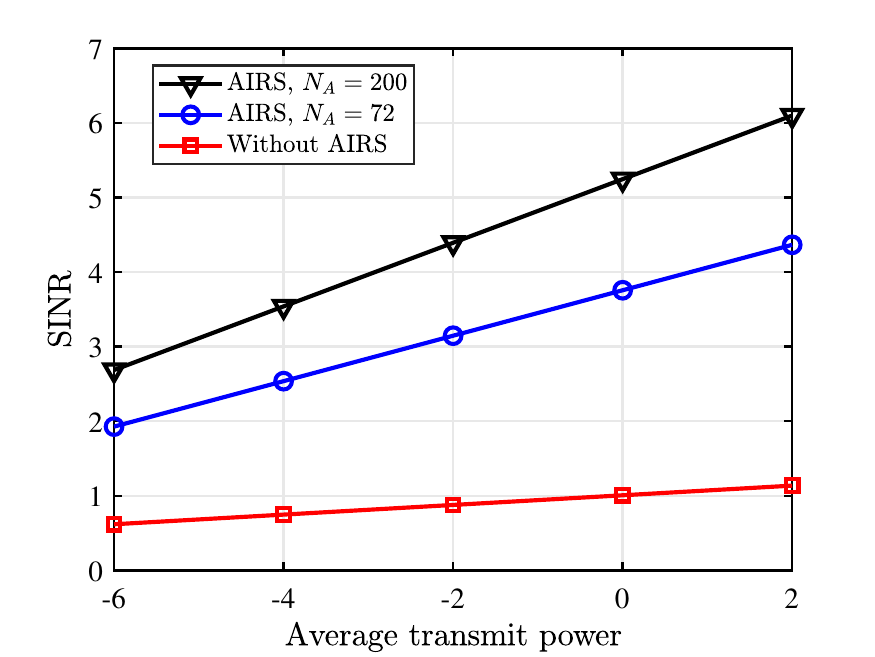} \\
		\caption{Impact of AIRS on Computation Offloading against DISCO jamming attacks.}\label{DDADSO against jammer}
	\end{center}
\end{figure}
\subsubsection{Effectiveness of AIRS-aided offloading in mitigating DISCO jamming}
Besides, Fig. \ref{DDADSO against jammer} illustrates the variation of received SINR with respect to transmit power under DISCO jamming attacks. We can see that the proposed AIRS-enhanced anti-jamming achieves a higher SINR than that of the DIRS-jammed MEC system without AIRS. For instance, when the transmit power $P$ = -2 dB and employing 72 AIRS reflecting elements, the SINR value of the AIRS-aided MEC system is 4.3933, representing a 258.90\% increase compared to the SINR value of the DIRS-jammed MEC system without AIRS. Furthermore, expanding the number of DIRS reflecting elements from 72 to 200 also leads to an even greater improvement, increasing by 39.61\%. These results demonstrate that AIRS effectively mitigates the DISCO jamming, thereby validating the necessity and effectiveness of incorporating AIRS phase shift optimization into the design of the proposed DDADSO scheme.

\subsection{Impact of AIRS Deployment on Computation Offloading}

\begin{figure}[t]
	\vspace{-0.3cm}
	\setlength{\abovecaptionskip}{0cm}
	\setlength{\belowcaptionskip}{-0.5cm}
	\begin{center}
		\includegraphics[width= 2.2 in]{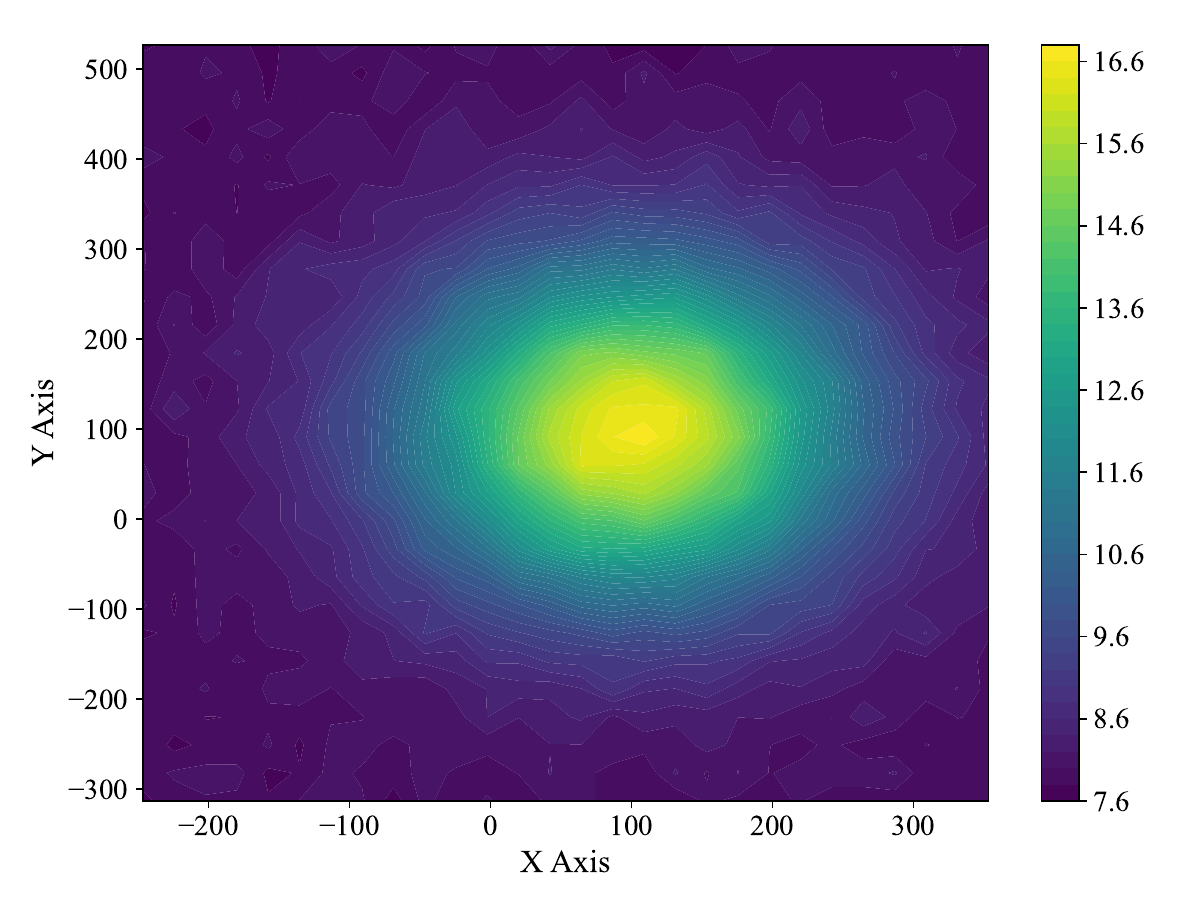} \\{\footnotesize (a) Different $x_A$ and $y_A$}\\
        \includegraphics[width= 2.2 in]{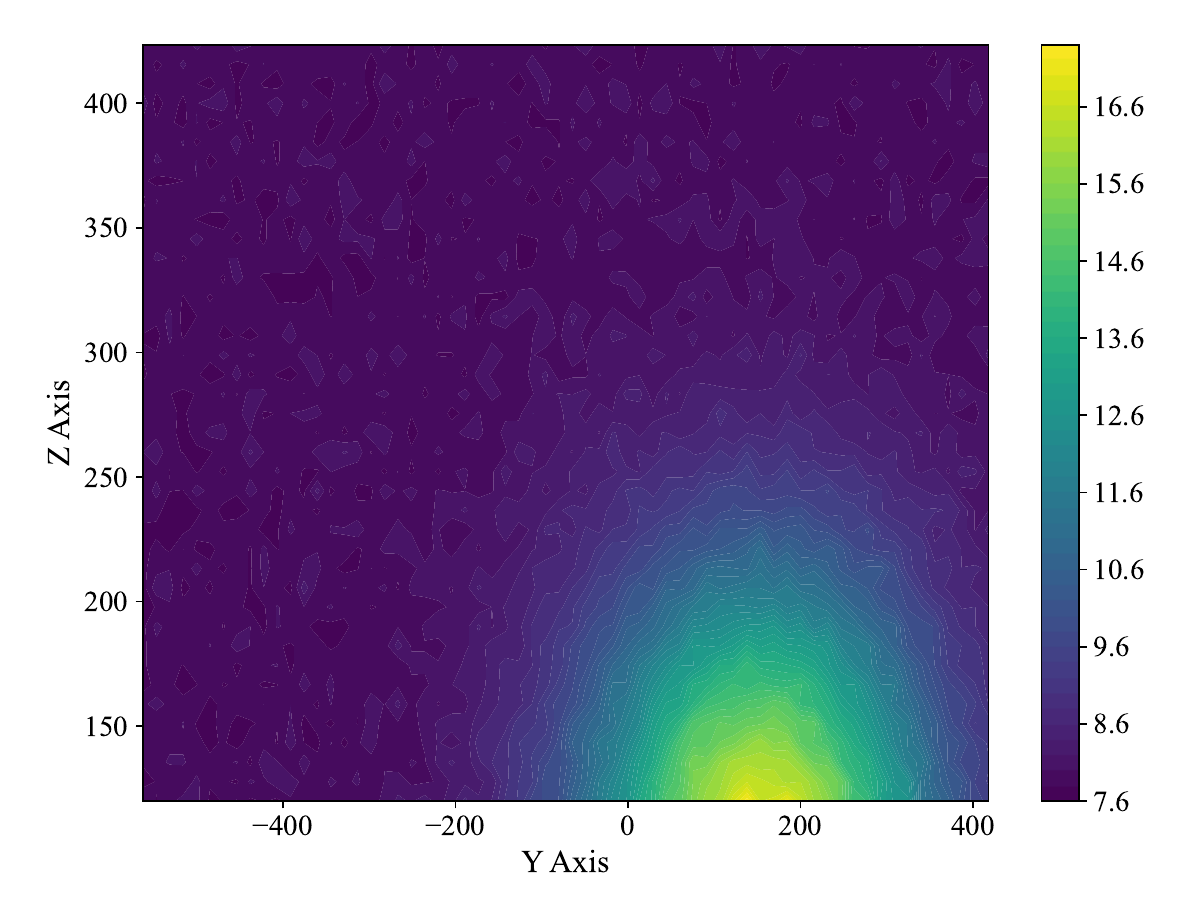} \\{\footnotesize (b) Different $y_A$ and $z_A$}\\
		\caption{The variation in offloading utility with different AIRS deployment against DISCO jamming.}\label{deployment}
	\end{center}
\end{figure}
To investigate the offloading performance of AIRS deployment against DISCO jamming, Fig.~\ref{deployment} presents utility heat maps for different AIRS positions. The upper subplot shows utility variation in the horizontal plane $(x_c, y_c)$, and the lower subplot illustrates its distribution in the vertical plane $(y_c, z_c)$. The offloading utility exhibits obvious spatial dependence with clear high-utility regions in both planes. In the horizontal plane, a centralized utility peak reveals that appropriate AIRS hovering positions significantly improve anti-jamming offloading performance. In the vertical plane, utility is concentrated at specific altitudes and Y-axis positions, indicating high sensitivity to deployment height and lateral location. These results differ from conventional AIRS-assisted communication studies \cite{Cheng2022RotationLocation}, highlighting the importance of AIRS deployment in suppressing DIRS interference and optimizing offloading. The irregular and nonlinear utility patterns highlight the necessity of dynamic AIRS deployment optimization within the proposed DDADSO scheme.

\subsection{Performance of Different Computation Offloading Schemes}
\begin{figure}[t]
	\vspace{-0.5cm}
	\setlength{\abovecaptionskip}{0cm}
	\setlength{\belowcaptionskip}{0cm}
	\begin{center}
		\includegraphics[width= 2.9 in]{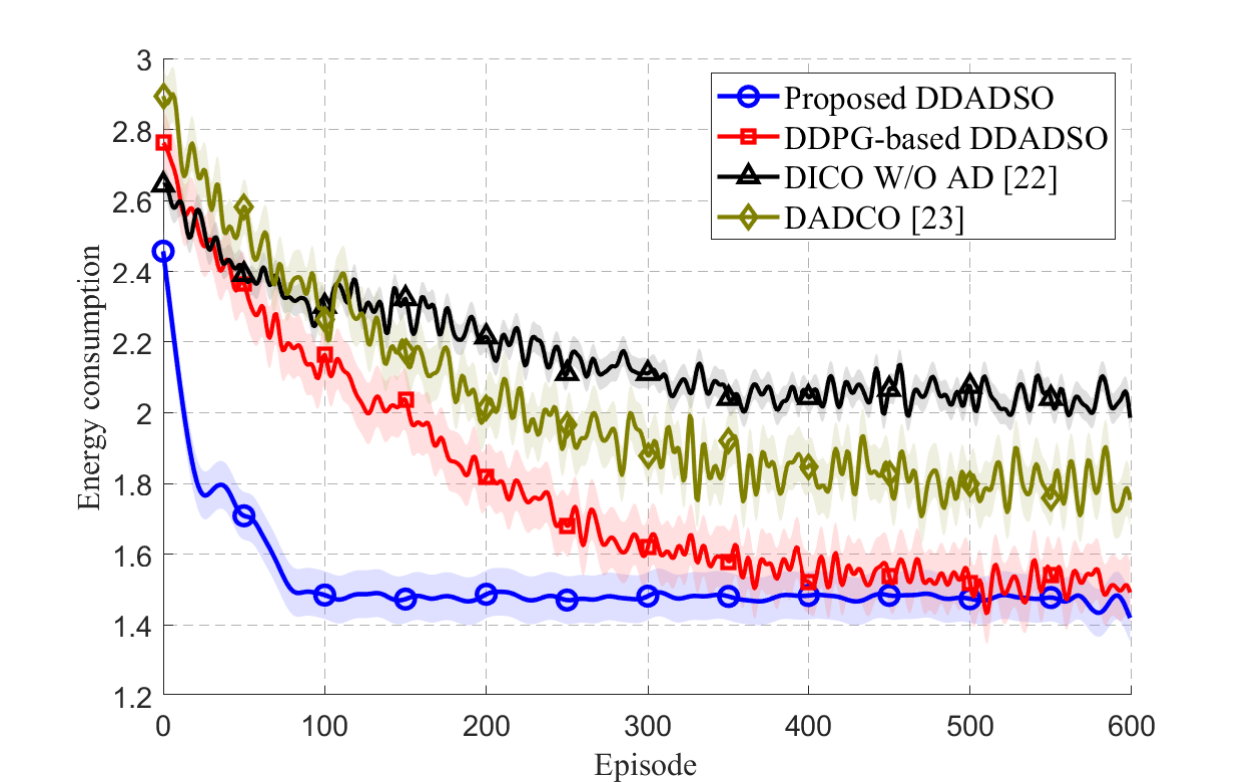} \\{\footnotesize (a) Energy consumption}\\
        \includegraphics[width= 2.9 in]{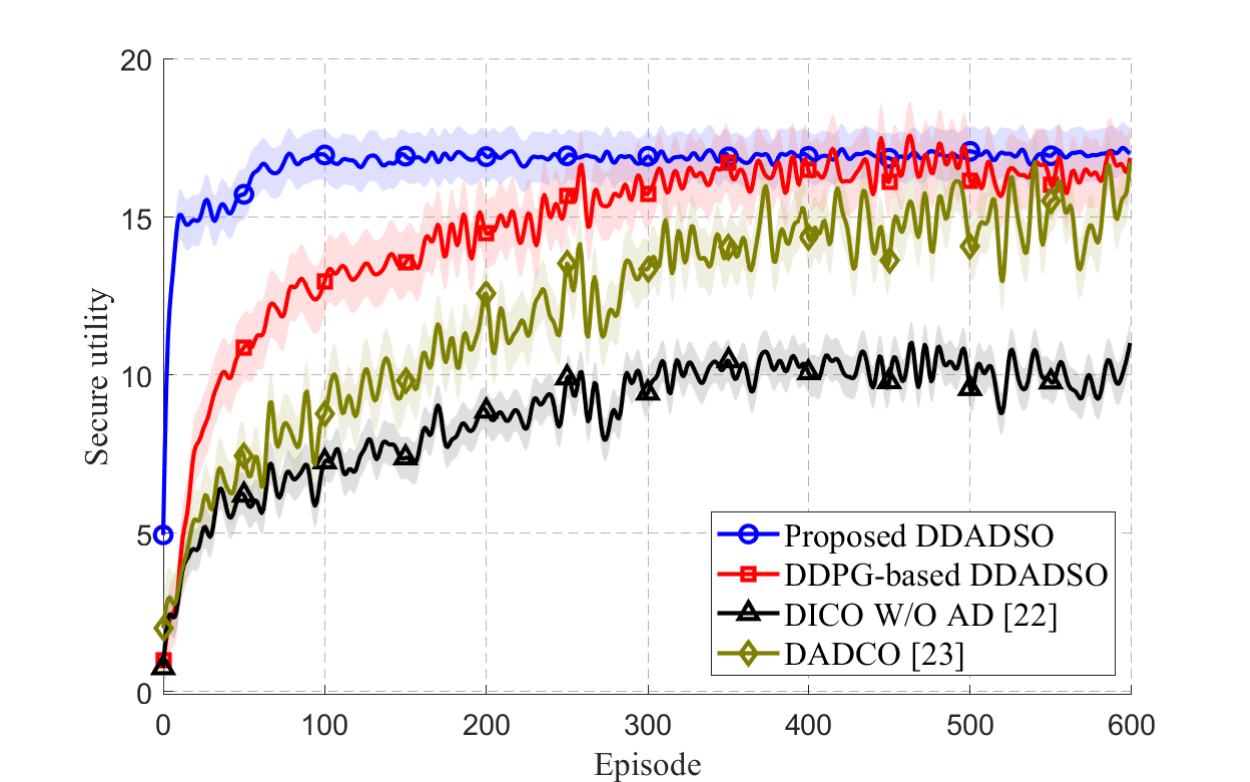} \\{\footnotesize (b) Secure utility}\\
		\caption{Convergence performance of the proposed DDADSO scheme compared to benchmark over 600 episodes. }\label{Convergence}
	\end{center}
\end{figure}
\subsubsection{Convergence performance of the DDADSO scheme and benchmark schemes}
Fig. \ref{Convergence} illustrates the performance advantages of the DDADSO scheme over 600 training episodes. As shown in Fig.~\ref{Convergence}(a), DDADSO outperforms DICO W/O AD and DADCO in energy consumption, achieving 25.27\% and 15.25\% energy reduction, respectively. In Fig.~\ref{Convergence}(b), the proposed DDADSO scheme gains a faster convergence speed and a similar steady-state utility compared to the DDPG-based DDADSO scheme. This enhanced performance is attributed to leveraging a dual-agent DRL algorithm that operates across different timescales and designing the encoder of the VAE. Additionally, AIRS deployment optimization enables DDADSO to achieve 75.50\% higher secure utility than DICO W/O AD with better convergence. To sum up, the proposed DDADSO scheme introduces the dual-agent DRL algorithm that combines the DQN and VAE-enhanced TD3 algorithms to determine the optimal AIRS deployment-aided secure computation offloading strategy.

\subsubsection{Impact of the number of AIRS reflecting elements on the different schemes}
\begin{figure}[t]
	\vspace{-0.5cm}
	\setlength{\abovecaptionskip}{0cm}
	\setlength{\belowcaptionskip}{0cm}
	\begin{center}
		
		\includegraphics[width= 2.4 in]{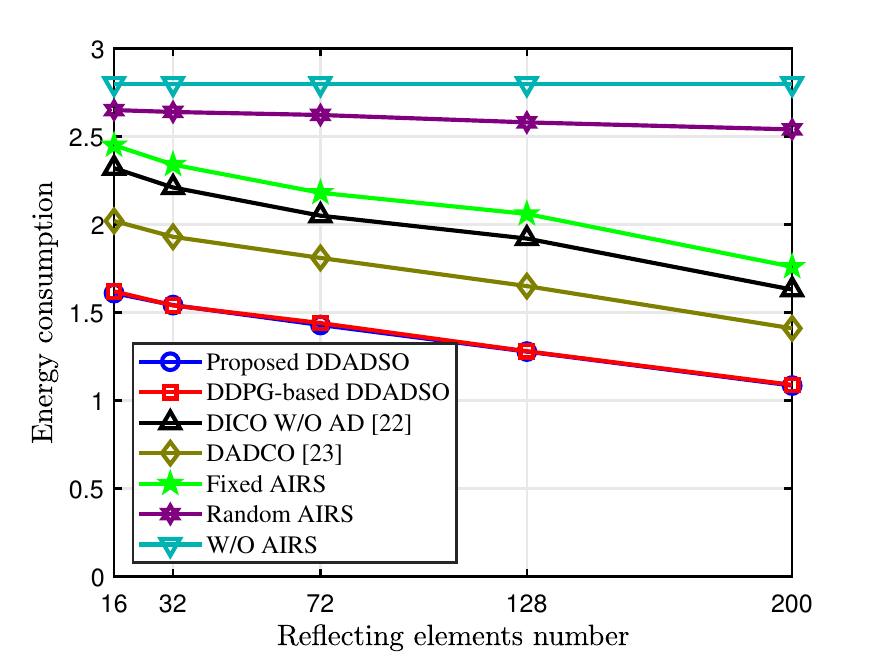} \\{\footnotesize (a) Energy consumption}\\
        \includegraphics[width= 2.4 in]{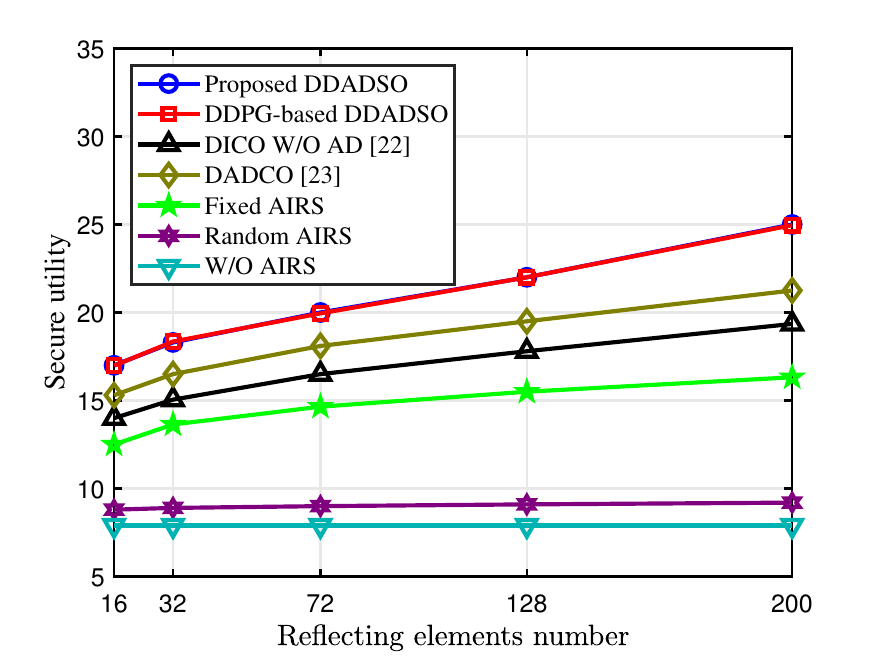} \\{\footnotesize (b) Secure utility}\\
		\caption{Impact of the number of reflecting elements on the system performance of different schemes.}\label{Anumber}
	\end{center}
\end{figure}

In Fig.~\ref{Anumber}, we evaluate the average performance of different IRS-based computation offloading schemes under varying numbers of reflecting elements. As shown in Fig.~\ref{Anumber}(a), the energy consumption of DDADSO and three benchmark schemes decreases as the number of AIRS reflecting elements increases from 16 to 200. In terms of the offloading utility in Fig.~\ref{Anumber}(b), DDADSO achieves more significant gains than the W/O AIRS condition, especially with more AIRS elements. Specifically, as AIRS reflecting elements increase, DDADSO, DICO W/O AD, DADCO, Fixed AIRS, and Random AIRS achieve improvements of 47.00\%, 38.89\%, 33.00\%, 30.56\%, and 4.55\%, respectively. This indicates the advantage of optimizing phase shift design and IRS deployment. Additionally, we observe that an increased number of reflecting elements results in a greater gain in beamforming, which minimizes the intensity of DISCO jamming.

\begin{figure}[t]
	\vspace{-0.5cm}
	\setlength{\abovecaptionskip}{0cm}
	\setlength{\belowcaptionskip}{0cm}
	\begin{center}
		
		\includegraphics[width= 2.7 in]{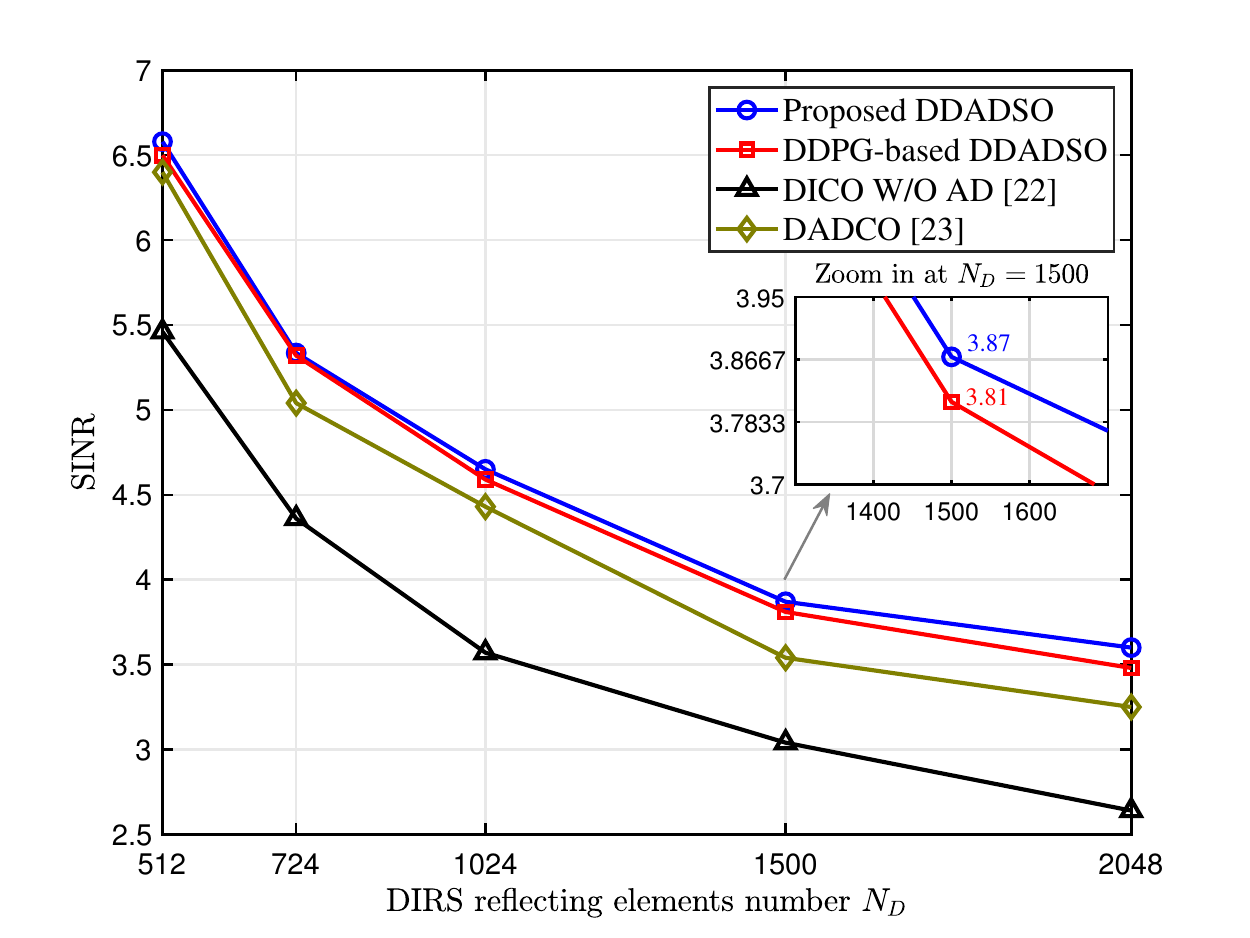} \\{\footnotesize (a) SINR}\\
        \includegraphics[width= 2.7 in]{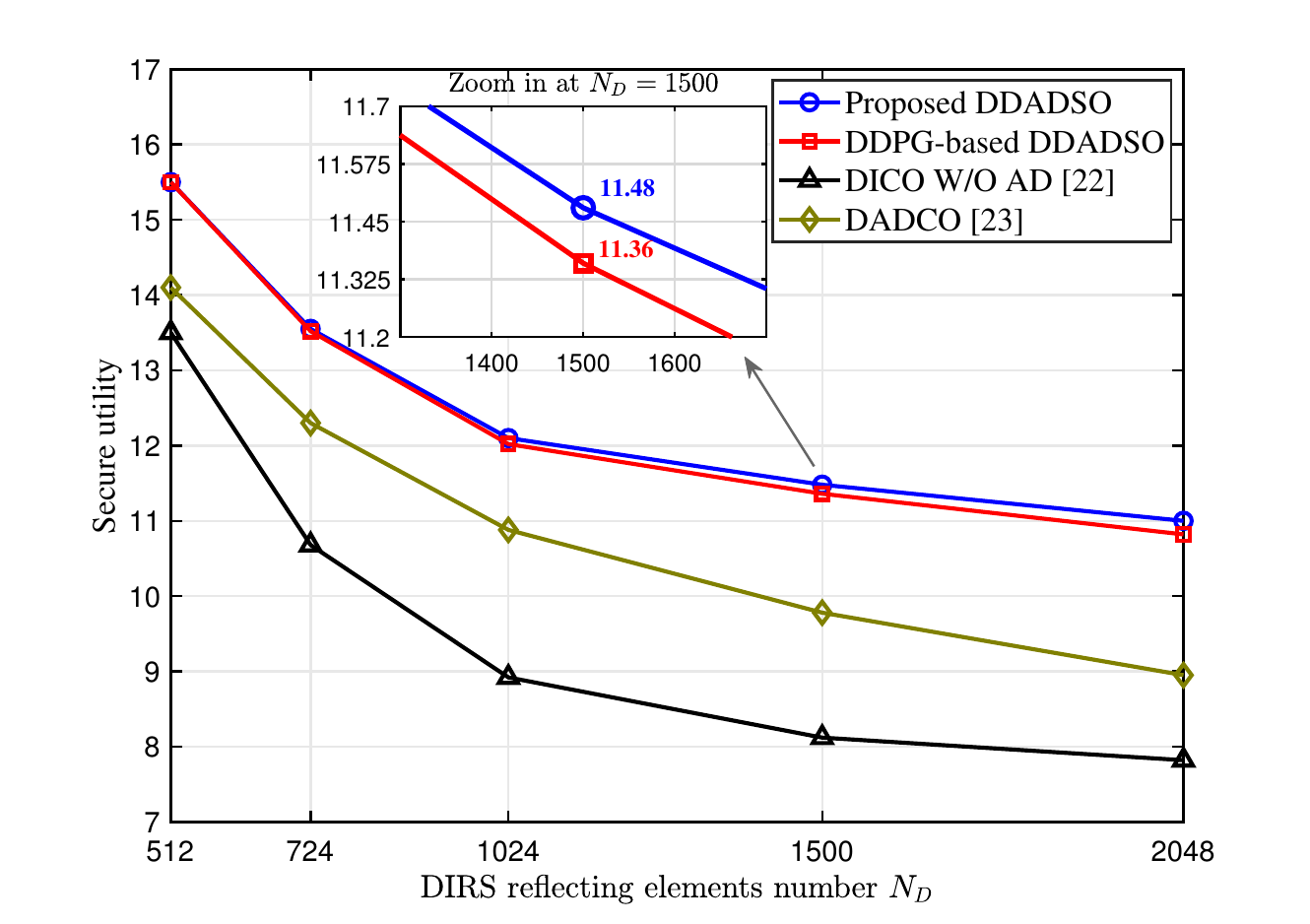} \\{\footnotesize (b) Secure utility}\\
		\caption{Impact of the number of DIRS reflecting elements  $N_D$ on the system performance of different schemes.}\label{DIRS_Number}
	\end{center}
\end{figure}
\subsubsection{Impact of the number of DIRS reflecting elements on the different schemes}
Fig.~\ref{DIRS_Number} illustrates the impact of increasing the number of DIRS reflecting elements on the offloading performance of various IRS-based computation offloading schemes. In Fig.~\ref{DIRS_Number}(a) and Fig.~\ref{DIRS_Number}(b), as the number of DIRS elements increases from 512 to 2048, all schemes exhibit performance degradation due to stronger DISCO jamming. However, the proposed DDADSO scheme maintains a larger utility than other schemes, and this difference becomes more apparent as the number of DIRS reflecting elements increases. In particular, when the number of DIRS elements is larger, the utility value of the proposed DDADSO scheme is notably bigger than that of the DDPG-based DDADSO scheme. Overall, these findings confirm that the proposed DDADSO scheme exhibits excellent robustness against intense DISCO jamming and leverages an improved algorithm to enhance system performance in such environments.

\begin{table*}[!t] 
\caption{Computational Overhead Comparison of Different Schemes}
\centering
\renewcommand{\arraystretch}{0.9} 
\resizebox{\textwidth}{!}{ 
\begin{tabular}{ccccc}
\toprule
\textbf{Scheme} & \textbf{\makecell{MEC Server \\ Memory (MB)}} & \textbf{\makecell{Computational Complexity}} & \textbf{\makecell{Policy Selection \\ Time (ms)}} & \textbf{\makecell{Convergence Time \\ (Episode)}} \\
\midrule
Proposed DDADSO  & 1256  & $\mathcal{O}\!\left(J\left(S_{in}n_z+n_zS_z+n_z^2+n_{\mu}^2+n_Q^2\right)+n_{DQN}^2\right)$  & 1.42 & 90 \\
DDPG-based DDADSO & 924  & $\mathcal{O}\left( J\left( n_{\mu}^{2}+n_{Q}^{2} \right) +n_{DQN}^2 \right)$ & 1.31 & 443 \\
DICO W/O AD [22] & 768 & $\mathcal{O}\left(  n_{\mu}^{2}+n_{Q}^{2}  \right)$ & 0.82 & 450  \\
DADCO [23] & 818 & $\mathcal{O}\left(  n_{\mu}^{2}+n_{Q}^{2}  \right)$ & 0.83 & 480  \\
\bottomrule
\end{tabular}
}
\label{Overhead}
\end{table*}

\begin{figure}[!t]
	\setlength{\abovecaptionskip}{0cm}
	\setlength{\belowcaptionskip}{0cm}
	\begin{center}
		\includegraphics[width= 2.7 in]{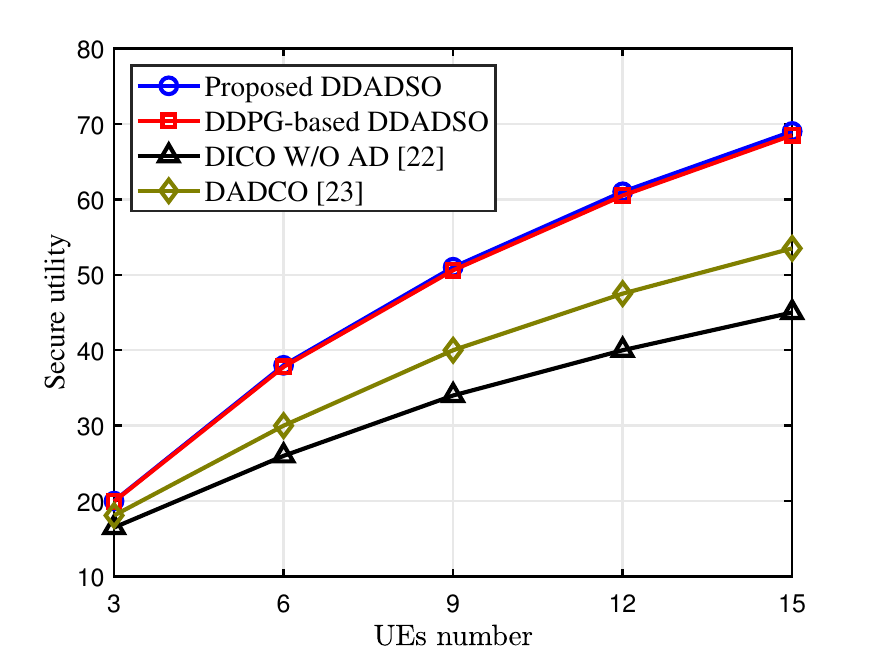}\\
		\caption{System performance with a growing number of UEs.}
\label{UEs}
	\end{center}
\end{figure}
\subsubsection{Impact of the number of UEs on the different schemes}
To verify DDADSO’s scalability to dynamic UE counts, we evaluate performance as $K$ increases from 3 to 15. The $K$ UEs are randomly distributed in a square serving area of $40\times 40\ \text{m}^2$ \cite{hu2021reconfigurable}. As shown in Fig. \ref{UEs}, when $K$ increases from 3 to 15, the proposed DDADSO obtains a significant performance gain of 2.45 times in terms of secure utility, while the benchmark schemes only exhibit relatively slow growth. This resilience stems from the VAE design and dual-agent collaboration, where the long-timescale AIRS deployment optimizes coverage to accommodate more UEs, while short-timescale phase shifts and offloading optimization balances each user's offloading. This result confirms that DDADSO exhibits excellent scalability to dynamically increasing UEs, validating its applicability to the dynamic number of UEs.

As shown in Table \ref{Overhead}, we compare the computational overhead of the proposed DDADSO scheme with three benchmarks on a Windows 10 platform equipped with a 2.9 GHz processor and 16 GB of RAM. The DDADSO scheme uses 1,256 MB of memory, which is more than the benchmarks, but it can still be supported by modern MEC servers \cite{Muramatsu202440}. Despite the increase in policy selection time due to its computational complexity, DDADSO converges to the optimal policy in only 90 episodes. This is about 79.68\%, 80.00\%, and 81.25\% faster than DDPG-based DDADSO, DICO W/O AD, and DADCO, respectively. This performance improvement demonstrates that the modest increase in computational complexity incurred by DDADSO is a worthwhile trade-off that significantly accelerates convergence speed and strengthens exploration efficiency.

\section{Conclusion}\label{Conclusion}
In this paper, we have addressed the novel DISCO jamming attack by designing an AIRS-aided secure computation offloading framework, which leverages adaptive deployment and phase shift adjustment to enhance anti-jamming performance. Furthermore, we have decomposed the problem into a two-timescale framework to align with the distinct update frequencies of long-timescale AIRS deployment and short-timescale AIRS-aided offloading policies, reducing redundant computations and improving learning efficiency. Then, we have modeled the AIRS deployment-aided secure computation offloading process as an MDP and proposed a dual-agent DRL algorithm to learn an optimal AIRS-aided secure computation offloading strategy through interaction with the environment and accumulated experiences. In the DDADSO scheme, the DQN-based agent optimizes the AIRS deployment on a long timescale to enhance anti-jamming capability, while the TD3-VAE-based agent optimizes the offloading ratios and phase shifts on a short timescale to adapt to the DIRS-jammed channel conditions. The VAE module compresses high-dimensional states to reduce computational complexity and alleviate learning non-stationarity. Finally, simulation results have shown that the proposed DDADSO scheme outperforms existing benchmarks in terms of the secure utility. In particular, the DDADSO scheme has achieved stable convergence and improves system utility against the DISCO jamming attacks.

\begin{spacing}{1.0}
\balance
\bibliography{reference}
\bibliographystyle{ieeetr}
\end{spacing}

\end{document}